\documentclass[sigconf]{acmart}

\AtBeginDocument{%
  \providecommand\BibTeX{{%
    \normalfont B\kern-0.5em{\scshape i\kern-0.25em b}\kern-0.8em\TeX}}}

\copyrightyear{2025}
\acmYear{2025}
\makeatletter
\def\@ACM@copyright@check@cc{}
\makeatother
\setcopyright{cc}
\setcctype{by}
\acmConference[UIST '25]{The 38th Annual ACM Symposium on User Interface
Software and Technology}{September 28-October 1, 2025}{Busan, Republic of
Korea}
\acmBooktitle{The 38th Annual ACM Symposium on User Interface Software and
Technology (UIST '25), September 28-October 1, 2025, Busan, Republic of
Korea}\acmDOI{10.1145/3746059.3747622}
\acmISBN{979-8-4007-2037-6/2025/09}

\usepackage{enumitem}
\usepackage{subcaption}
\usepackage{caption}
\usepackage{flushend}
\usepackage{xspace}
\newcommand{\alphaval}[2]{{\small $p\,#1\,#2$}}
\newcommand{\statsum}[3]{{\small $M=#1\,#3$, $SD=#2\,#3$}}
\newcommand{\statsd}[2]{{\small $SD=#1\,#2$}}
\newcommand{\friedman}[4]{{\small $\chi^{2}(#1)=#2$, \alphaval{#3}{#4}}}

\newcommand{\ttest}[4]{{\small $t(#1)=#2$, \alphaval{#3}{#4}}}
\newcommand{\mann}[3]{{\small $W=#1$, \alphaval{#2}{#3}}}

\newcommand{\notransfer}[1]{\emph{No-Transfer#1}}
\newcommand{\crossusers}[1]{\emph{Cross-Users#1}}
\newcommand{\crossthemes}[1]{\emph{Cross-Themes#1}}
\newcommand{\metapo}[1]{\emph{Meta-PO#1}}

\newcommand{\anova}[6]{{\small [$F_{#1,#2}$\,$=$\,$#3$, $p$\,$#4$\,$#5$]}}
\newcommand{\pvall}[2]{{\small $p\,#1\,#2$}} 

\newcommand \change[1]{{\textcolor{black}{#1}}}
\newcommand\delete[1]{}
\definecolor{blue}{rgb}     {0,0.0,1.0}

\newcommand{\method}{Meta-PO\xspace}
\newcommand{\methodnospace}{\method}

\author{Zhipeng Li}
\affiliation{%
  \institution{Department of Computer Science}
  \institution{ETH Z\"urich}
  \country{Z\"urich, Switzerland}
}
\email{zhipeng.li@inf.ethz.ch}

\author{Yi-Chi Liao}
\affiliation{%
  \institution{Department of Computer Science}
  \institution{ETH Z\"urich}
  \country{Z\"urich, Switzerland}
}
\email{yichi.liao@inf.ethz.ch}

\author{Christian Holz}
\affiliation{%
  \institution{Department of Computer Science}
  \institution{ETH Z\"urich}
  \country{Z\"urich, Switzerland}
}
\email{christian.holz@inf.ethz.ch}

\begin{document}

\title[Efficient Visual Appearance Optimization by Learning from Prior Preferences]{Efficient Visual Appearance Optimization \\ by Learning from Prior Preferences}

\begin{abstract}


Adjusting visual parameters such as brightness and contrast is common in our everyday experiences. 
Finding the optimal parameter setting is challenging due to the large search space and the lack of an explicit objective function, leaving users to rely solely on their implicit preferences.
Prior work has explored Preferential Bayesian Optimization (PBO) to address this challenge, involving users to iteratively select preferred designs from candidate sets.
However, PBO often requires many rounds of preference comparisons, making it more suitable for designers than everyday end-users.
We propose Meta-PO, a novel method that integrates PBO with meta-learning to improve sample efficiency.
Specifically, Meta-PO infers prior users' preferences and stores them as models, which are leveraged to intelligently suggest design candidates for the new users, enabling faster convergence and more personalized results.
An experimental evaluation of our method for appearance design tasks on 2D and 3D content showed that 
participants achieved satisfactory appearance in 5.86 iterations using Meta-PO when participants shared similar goals with a population (e.g., tuning for a ``warm'' look) and in 8 iterations even generalizes across divergent goals (e.g., from ``vintage'', ``warm'', to ``holiday'').
Meta-PO makes personalized visual optimization more applicable to end-users through a generalizable, more efficient optimization conditioned on preferences, with the potential to scale interface personalization more broadly.

\end{abstract}

\ccsdesc[500]{Computing methodologies~Machine learning}
\ccsdesc[500]{Computing methodologies~Graphics systems and interfaces}
\ccsdesc[500]{Human-centered computing}

\keywords{Human-in-the-loop optimization, visual design, preferential optimization, preferential modeling, Bayesian optimization, meta-learning, meta-Bayesian optimization}

\begin{teaserfigure}
    \centering%
    \includegraphics[width=\linewidth]{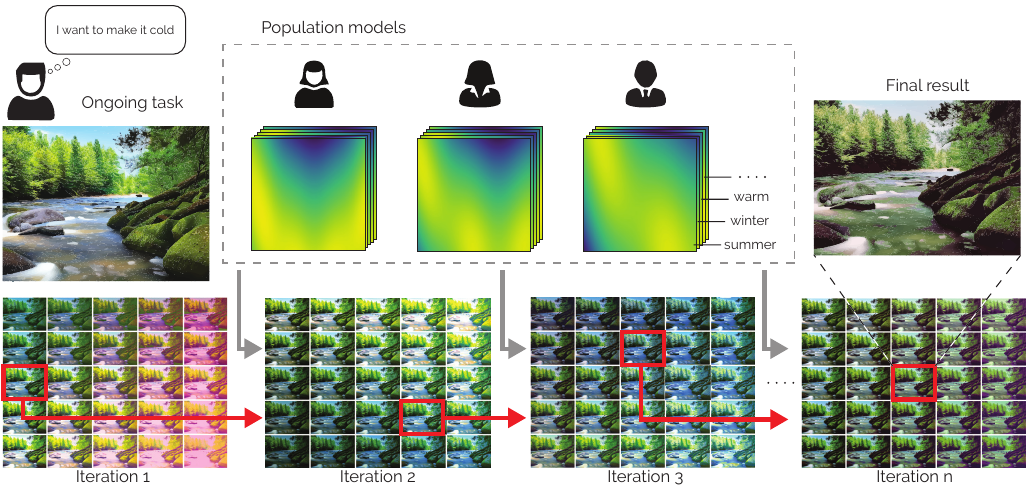}%
    \vspace{-2mm}%
    \caption{
        We introduce \emph{\method}, a computational method that models users' implicit preferences of visual appearances given a theme (here "cold") as they iteratively select from a sampled set of candidates presented in a 2D gallery. 
        \method integrates Preferential Bayesian Optimization with meta-learning to increase optimization efficiency by leveraging population models derived from \emph{prior optimization experiences} --- across users and themes.
        \method can thus produce more satisfying results in fewer iterations.
    }
    \label{fig:teaser}
\end{teaserfigure}


\maketitle

\section{Introduction}

Adjusting visual parameters is a common part of everyday life.
Most people tune image appearance by adjusting brightness, contrast, or applying filters, before sharing photos on social media or for professional use~\cite{chen2019association}.
However, this task is challenging for two reasons.
First, the search space is overwhelmingly large.
There are 10 to 15 common parameters that affect image appearance, e.g., exposure, saturation, hue, and sharpness.
Even with just 10 parameters discretized into 10 levels each, the number of combinations reaches $10^{10}$, making a thorough manual search impractical.
Second, visual design lacks an explicit objective function.
The ``goodness'' of a design is subjective, governed by the user's implicit preferences. Thus, end-users rarely try to find their ideal appearance from scratch, and instead rely on pre-defined filters \cite{manikonda2014analyzing}. 
Even professional designers, though able to spend more time exploring, often settle for a result that is ``good enough'' without a clear sense of optimality~\cite{lawson2006designers, cross2004expertise}.

\change{To address this challenge, prior works introduced Preferential Bayesian Optimization (PBO) as a principal approach, along with applying existing filters and tuning parameters.}
PBO builds on Bayesian Optimization (BO), a probabilistic framework for solving black-box functions whose internal structure is inaccessible and must be learned through iterative evaluations \cite{frazier2018tutorial, mockus1975bayes}.
BO is widely used for human-in-the-loop tasks due to its robustness, generalizability, and ability to handle uncertainty~\cite{shahriari2015taking, krause2011contextual}.
PBO extends BO to settings where the values of the objective functions are \textit{not directly observable}, relying instead on inferred user preferences through comparative feedback.
Practically, PBO operates via selection-based interaction: users choose the most preferred design among a presented set.
By observing these selections, the optimizer gradually learns a model of the user's implicit preferences, eventually converging on the most preferred design.
\change{With this mechanism, PBO enables users to navigate design spaces where manual tuning is impractical due to high dimensionality or a lack of pre-defined filters. 
This capability has demonstrated its effectiveness and efficiency in various domains, including fabrication \cite{deneault2025preferential}, font design \cite{tatsukawa2025fontcraft}, and color selection \cite{iwai2025constrained}.}

Prior work has proposed various interactions to improve sample efficiency in image enhancement using PBO. 
For example, linear search spaces~\cite{koyama2017sequential} and gallery-based interfaces~\cite{koyama2020sequential} have been used to elicit preferences beyond the typical pairwise comparisons~\cite{gonzalez2017preferential}. 
Among these, the gallery-based interaction (illustrated in \autoref{fig:teaser}) achieves the highest efficiency, as it enables exploration of a 2D search plane in one iteration.
\change{These studies also showed that PBO allows users to produce images that are more preferred than those created using pre-defined filters in Photoshop and Lightroom.}
Despite these advances, PBO still requires 10 to 15 iterations to converge~\cite{koyama2020sequential}, potentially limiting its practicality in everyday applications.
In this paper, we aim to enable PBO to \emph{converge even more efficiently toward a user's implicit preference} by leveraging prior users' preference data, making it suitable for end-users in everyday visual design tasks.
This idea is rooted in transfer learning (adapting to new but related tasks) and meta-learning (learning to learn).
While both have been widely studied in machine learning and optimization \cite{vilalta2002perspective, weiss2016survey, bai2023transfer}, only recently have they been explored in human-in-the-loop tasks \cite{liao2024meta}.
To our knowledge, no prior work has applied meta-learning to preferential optimization, where preferences are inferred via comparisons.

We propose a novel computational method, \method (short for ``\textbf{Meta}-learning for \textbf{P}referential Bayesian \textbf{O}ptimization''), to enable efficient visual appearance optimization. 
The core principle of \method is to transfer prior users' preference models (learned from past optimization processes) on a specific theme (e.g., mood, theme, or filter) to accelerate the optimization process for new users with a similar target theme.
For example, if prior users have previously completed PBO sessions tuning images to offer a ``warm'' mood, we construct separate models from their data. 
These models can then be leveraged to boost optimization efficiency for a new user aiming to achieve a similar ``warm'' visual effect.
Our assumption is that while individual interpretations of ``warm'' vary slightly and target images differ, there are shared \textbf{across-users} preference patterns, which can support the future optimization.
Furthermore, we extend this concept to a more challenging setting: \textbf{cross-theme generalization}. 
Suppose we have accumulated preference optimization data across diverse moods, themes, or filters (e.g., warm, winter, summer, happy, sad, vintage, autumn, and more). 
When a new user aims to adjust images for an unseen goal (e.g., ``cold''), our method can still utilize relevant prior data, as shown in~\autoref{fig:teaser}.
\method achieves this by automatically identifying the relevance of prior models to adjust their influence level on the current optimization, allowing knowledge transfer from related themes while minimizing interference from unrelated ones. 
Moreover, \method has an internal mechanism where the guidance of prior models is stronger in the initial phase of an optimization, and gradually, the outcome will be tailored only based on the observations from the new user.

Several computational elements are integrated to build \methodnospace. 
Our interaction is built on the Sequential-Gallery \cite{koyama2020sequential}, where users are presented with a 2D plane of visual design options sampled from a high-dimensional design space, allowing them to select the most preferred design (the interaction is illustrated in~\autoref{fig:teaser}).
We incorporate an aggregation-based meta-learning mechanism: a set of prior preferential models is constructed from previously collected user data, and these models collectively guide the ongoing optimization.
Each model's influence is determined by dynamically assigned weights, with higher weights indicating stronger influence.
Specifically, we adopt a variant of the Transfer Acquisition Function (TAF-R) \cite{wistuba2018scalable}, which computes similarity between the current user's preferences and those of prior users, enabling selective and adaptive knowledge transfer.
To construct the 2D gallery \cite{koyama2020sequential}, we require at least three design points. 
As in the original method, we select the current best design (the user's most preferred so far) as the first point. We then compute the second point using TAF-R, leveraging meta-learned knowledge to guide the search.
Our final novelty lies in the selection of the third design point. We aim to, again, leverage prior knowledge without proposing a point too similar to the second. 
We combine TAF-R with the 2-step Bayesian optimization method \cite{wu2019practical}, which identifies a candidate likely to provide significant improvement if the current most promising design were to be evaluated. This ensures diversity and exploration in the 2D gallery presented to the user.

We evaluate \methodnospace’s \textbf{cross-user transfer} and \textbf{cross-theme generalization} through user studies, demonstrating improved efficiency on two distinct applications: image enhancement and virtual reality lighting.
With prior optimization experience on the same theme, participants adjusted images to their desired visual theme in 5.86 (\statsd{1.20}) iterations, compared to 9.54 (\statsd{2.19}) without leveraging previous experience. 
In cross-theme scenarios, leveraging optimization experience from different themes, they required 7.41 (\statsd{1.28}) iterations, also significantly fewer than when no prior experience was utilized.
Participants also reported that \method enables more effective exploration and more satisfying outcomes. 
In virtual lighting design, \method users achieved satisfactory results in fewer iterations (\statsum{4.05}{1.39}{}) than those without knowledge transfer (\statsum{6.85}{1.48}{}), demonstrating its effectiveness in this application as well.
To summarize, we make the following contributions:

\begin{itemize}[leftmargin=*]
    \item We introduce \methodnospace, a novel method incorporating several computational elements to enable efficient preference-driven optimization, allowing both end-users and designers to identify their most preferred designs within just a few selections.
    \item We demonstrate the effectiveness of \method in achieving cross-user and cross-theme adaptation across both 2D and 3D visual design tasks, showcasing its generalizability and efficiency in real-world scenarios.
\end{itemize}

\section{Related work}

\subsection{Preferential Bayesian Optimization}
\label{section: related_work_BO}

Optimizing human-centered interfaces is challenging because the only information we have regarding the quality of the design is acquired via user evaluation. 
Bayesian optimization (BO) has long been employed to support this process.
Formally, the optimization process seeks $x^{*} = \arg\max_{x \in \mathcal{X}} f(x),$ where \(x\) is the parameter vector, and \( f(\cdot) \) is the human related objective function. 
In each iteration, the optimizer suggests a parameter setting, and the user provides corresponding feedback. 
BO relies on two fundamental elements: (1) a \textit{surrogate model}, typically a Gaussian Process (GP), which approximates the objective function, and (2) an \textit{acquisition function}, which selects the next evaluation point.  
\change{These two components enable BO to effectively manage the uncertainty in human-related objective functions, achieving a balance between exploration and exploitation under limited evaluation budgets. 
This capability makes BO particularly well-suited for human-in-the-loop tasks, where evaluations involving human feedback tend to be expensive and time-consuming~\cite{williamson2022bayesian, chan2022investigating, liao2023interaction}.
Leveraging these strengths,}
BO has found applications in a wide range of tasks because of its robust performance~\cite{frazier2018tutorial, wang2023recent, shahriari2015taking}, including input techniques~\cite{liao2024meta}, designing software interfaces~\cite{mo2024cooperative} and physical interactions~\cite{liao2023interaction, liao2020button}, refining haptic interfaces~\cite{catkin2023preference}, game design \cite{10.1145/2858036.2858253}, \change{UI design in urban air mobility~\cite{meinhardt2025fly}}, and other human-involved interactions.


In the above examples, BO directly observes the measurable quality of the design. 
However, there is another type of task that is unrealistic to expect human users to provide precise quantitative assessments.
For example, asking users to tell the goodness of a parameter configuration for adjusting photos' presentation. 
The goodness function is implicit in the human mind, which can not be measured directly. 
Prior works addressed this challenge by Preferential Bayesian Optimization (PBO)~\cite{gonzalez2017preferential}.  
Instead of requiring numerical evaluations, PBO asks users to compare the results of two configurations (\(x_a\) and \(x_b\)) and express a preference (e.g., \( f(x_a) > f(x_b) \) or \( f(x_a) < f(x_b) \)).  
These comparisons are then modeled probabilistically using approaches such as the Bradley-Terry-Luce (BTL) model~\cite{koyama2017sequential, tsukida2011analyze}, allowing PBO to integrate preference-based feedback within the standard BO framework to suggest the next evaluation candidates. 
\change{This makes PBO achieve greater efficiency and robustness compared to standard BO that relies on direct numerical ratings, especially when optimizing subjective or implicit functions~\cite{gonzalez2017preferential, takeno2022preferential}.}
Prior works explored interactive approaches that allow the users to identify the most preferred design among a set of options to further improve the efficiency \cite{koyama2017sequential, yamamoto2022photographic}, which will be reviewed in Preliminaries.

Although PBO is effective, its initial exploration phase is inefficient due to the lack of prior knowledge, requiring extensive random sampling~\cite{wang2023recent}.  
However, visual appearance functions often exhibit similarities across tasks.  
Leveraging this structure, we propose using meta-learning to accelerate early exploration and improve efficiency in visual appearance design.

\subsection{Meta-learning for Bayesian optimization}
\label{section: related_work_meta_learing}

Meta-learning is a machine learning paradigm that aims to enable models to quickly adapt to new tasks by leveraging prior experience from similar tasks~\cite{schweighofer2003meta, wang2021meta}. 
It has effectively improved the adaptation efficiency for recognition tasks and reinforcement learning~\cite{duan2016rl, finn2017model, nichol2018reptile}. 
Recent research in BO has attempted to integrate the concept of meta-learning into black-box optimization problems. 
That is, improving optimization efficiency by incorporating knowledge from previous optimization processes for similar problems. 
While Meta-Bayesian optimization (Meta-BO) has been employed in hyperparameter tuning and various machine learning tasks~\cite{li2021population, wang2023recent, li2022transbo}, its application to human-computer interaction (HCI) problems has only recently been explored~\cite{liao2024meta}.

There are multiple approaches to realizing Meta-BO. 
A straightforward method is to integrate data from the current task into a surrogate model (typically a Gaussian Process) that has been trained on previous tasks, which provides strong prior knowledge~\cite{bardenet2013collaborative, swersky2013multi, yogatama2014efficient}. 
However, GPs do not scale well to large datasets due to their cubic computational complexity; that is, $O(n^3)$, where $n$ represents the total number of observations across all tasks.
Another approach involves replacing the surrogate model or acquisition function with deep neural networks to handle large datasets~\cite{volpp2019meta, wistuba2021few, hsieh2021reinforced}. 
However, these methods typically require substantial amounts of training data and generally lack explainability, which is challenging or even impractical when the data requires human evaluations.
\change{Recently, \citeauthor{liao2025continual} proposed CHiLO, which uses a BNN as a population model trained on samples from individual users' GP, reducing the need for large amounts of user data while preserving data quality~\cite{liao2025continual}.
However, directly applying this approach to preferential tasks may be problematic, as it assumes all users' functions are inherently similar and can be captured by a single model.
In reality, human preferences often vary widely across users and tasks, and this might enforce the shared model risks over-smoothing user-specific differences.}

Another approach is the weighted-sum method, which constructs individual GP models for each task and aggregates their acquisition functions based on weighted contributions~\cite{li2022transbo, wistuba2018scalable}. 
This approach is more practical compared to the other methods: it does not introduce cubically increasing complexity or assume a large amount of training data.
\citeauthor{liao2024meta} showed this method can efficiently adapt the input device to replace manual calibration~\cite{liao2024meta}.
\change{However, these methods have focused on optimizing measurable objectives, such as speed or accuracy. 
Extending it to preference-based tasks remains an open challenge for efficiently optimizing preference objectives.}
In this work, we took this approach but incorporated novel extensions for addressing the specific needs of preference-driven tasks.
PBO must account for individual user preferences, which can vary significantly and even conflict at times. 
Further, the number of suggested designs increases in the PBO, requiring additional mechanisms. 
We address this by using multi-step BO, which will be detailed in Preliminaries. 

\subsection{Model-based Optimization for Design}

Design is a core topic in HCI, with computational methods increasingly used to support designers and users in creating effective, user-friendly systems~\cite{huang2009challenges}.
A widely adopted approach is model-based optimization, which formulates design problems as mathematical optimization tasks and evaluates candidates using predictive models rather than direct human feedback~\cite{liao2023human}.
Unlike human-in-the-loop optimization methods, model-based approaches rely on surrogate models to assess design quality, enabling faster iteration and scalable design generation.
These methods have been successfully applied in domains where reliable physical models are available to guide design decisions.
For example, they have been used to optimize geometry for acoustic control~\cite{li2016acoustic}, rotational inertia~\cite{bacher2014spin}, and object balance~\cite{prevost2013make}, while respecting aesthetic and structural constraints.
In UI design, optimization has improved keyboard layouts~\cite{karrenbauer2014improvements} and enabled real-time layout refinement through predictive models~\cite{todi2016sketchplore, o2015designscape}.
In Mixed Reality (MR), where interfaces extend beyond traditional screens, optimization techniques help adapt UI layouts to real-world spatial contexts~\cite{li2024situationadapt, cheng2023interactionadapt}.
Despite their broad applicability, model-based methods face two key limitations.
First, they require task-specific models, which may be infeasible or unavailable for many human-centered tasks.
Second, user preferences are inherently subjective and vary widely across individuals.
While aesthetic models exist to capture broad stylistic attributes such as vibrancy, warmth, or minimalism~\cite{datta2006studying, deng2017image}, they fall short in modeling individual differences of preferences, ultimately limiting true personalization.

\section{Preliminaries}
\label{sec:preliminaries}
We now review the areas that laid the foundation for \method. 
First, we review the development and underlying method of Preferential Bayesian Optimization (PBO). 
We then review the interactive systems that extend PBO for interactive systems, particularly focusing on the Sequential Gallery Interaction~\cite{koyama2020sequential}. 
Furthermore, we review a specific meta-BO method, Transfer Acquisition Function (TAF)~\cite{wistuba2018scalable}, which is the underlying approach for our method. 
Finally, we review a practical approach for 2-step Bayesian optimization \cite{wu2019practical} that is a critical part of \method to construct a user search plane at each iteration.

\subsection{Modeling Human's Implicit Preference}


The goal of PBO is to model implicit human preferences and optimize design accordingly. 
\change{There are two approaches to eliciting these preferences. A standard and more robust method is to present users with a set of candidates and ask them to choose their preferred option. This comparative format is generally easier for users to engage with and provides richer, more informative feedback, from which PBO can infer preference models. Prior studies have shown this approach to be effective for preference-based optimization tasks~\cite{koyama2017sequential, koyama2020sequential}.}

\change{A simpler alternative is to ask users to provide scalar ratings indicating how much they like a given option. However, this method suffers from fundamental limitations. First, subjective ratings are often noisy and inconsistent: it is inherently difficult for users to translate their preferences into precise numerical values, and individuals tend to apply different rating scales. Moreover, as identified in prior work~\cite{liao2024practical}, scalar ratings are strongly biased by recent experiences, further reducing reliability.
Second, scalar-based feedback is less efficient. In line-based or gallery-based PBO, each iteration yields multiple pairwise comparisons, accelerating convergence. In contrast, scalar ratings produce only a single data point per iteration, offering less information and requiring significantly more rounds of interaction. For these reasons, scalar-based PBO is rarely used in practice, except in settings where displaying multiple design candidates is infeasible, such as in driving scenarios~\cite{10.1145/3706598.3714187}.
}
\delete{To achieve this, PBO presents a set of candidate options to human users, who then compare them and select their most preferred choice. }

\change{We adopt the typical PBO approach by presenting a set of candidates to users, who compare them and select their most preferred choice.}
The selection results can be mathematically modeled using stochastic models such as the Bradley-Terry-Luce (BTL) model~\cite{tsukida2011analyze}.
With the probabilistic representation of the selection results, PBO can fit a Gaussian Process (GP) with the standard Bayesian Optimization (BO) framework and decide the next set of candidate options with its acquisition function for the subsequent iterations. 
As more selection results are accumulated over time, the Gaussian Process progressively refines its model of the user's preferences.

The core difference between PBO and standard BO is that PBO has to translate the selection results into likelihoods. 
The BTL model, which has been widely used in previous PBO methods, is well-suited for handling such selection scenarios~\cite{tsukida2011analyze}.
Specifically, consider \( m \) design candidates presented to the user, denoted as \( \mathcal{P} = \{x_i\}_{i=1}^m \), from which the user selects \( x^{c} \) based on his preference. 
Then we collect the feedback in the following form:  

\begin{equation}
    x^{c} \succ \mathcal{P} \backslash \{x^{c}\}
\end{equation}

\noindent
where \( \succ \) indicates that the perceived goodness value of \( x^{c} \) is greater than that of any other candidates in $\mathcal{P}$.
The BTL model defines the likelihood of this outcome as:

\begin{equation}
    p(x_i \succ \mathcal{P} \backslash {x_{i}} \mid \{g_j\}_{j=1}^{m}) = \frac{exp(g_i)}{\Sigma_{j=1}^{m} exp(g_j)}
\end{equation}

\noindent
where \( g_j\) represents the users' implicit goodness function for design candidate $j$. 
This allows us to fit the hyperparameters $\theta$ of the Gaussian Process by maximizing the posterior distribution as such:

\begin{equation}
    \theta = \arg \max_{\theta} p(\theta \mid \mathcal{D}_k) = \arg \max_{\theta} \frac{p(\mathcal{D}_k \mid \theta)p(\theta)}{p(\mathcal{D}_k)}.
\end{equation}
\label{eq:modelupdate}

\noindent
Here, \(\mathcal{D}_k\) comprises all the prior user's preference responses until $k$-th iteration, and \( p(\mathcal{D}_k) \) is a normalizing constant.
\( p(\theta) \) is the prior distribution over \( \theta \), which is predefined, and $p(\mathcal{D}_k \mid \theta)$ is the posterior distribution after all the observations, which can be rewritten as 

\begin{equation}
    p(\mathcal{D}_k \mid \theta) = \int p(\mathcal{D}_k \mid g) p(g \mid \theta) dg
\end{equation}

\noindent
where \( p(g \mid \theta) \) is the prior probability defined by \( \theta \).  
Since \( \mathcal{D}_k \) consists of a series of selection results, we compute the likelihood of the full dataset as a product over all comparisons observed so far.  
Let each iteration \( i \) involve a user selecting the preferred design \( x^c_i \) from a candidate set \( \mathcal{P}_i \). Then:

\begin{equation}
    p(\mathcal{D}_k \mid g) = \prod \limits_{i=1}^{k}  p(x^{c}_i \succ \mathcal{P}_i \backslash \{x^{c}_i\} \mid g)
\end{equation}

\noindent
With this formulation, the hyperparameters \( \theta \) can be estimated using the series of preference-based selection results observed up to iteration \( k \).  
A detailed procedure for fitting the Gaussian Process using the BTL model can be found in \cite{koyama2017sequential}.




\subsection{Interactions with PBO}

With the implicit preference model, we still need to decide the next set of candidates and present them to users for selection.
Prior works have implemented interactive systems that allow users to easily explore a low-dimensional search space constructed from the preference model and select the most preferred design instance.
Sequential-Line adopts a linear search space, while Sequential-Gallery presents a 2D gallery of design candidates~\cite{koyama2017sequential, koyama2020sequential}.
\change{The gallery-based interaction interface has been shown to encourage more effective exploration and inspire users to try diverse options, resulting in a more engaging and creative experience compared to traditional filter-based methods~\cite{koyama2020sequential}.}

Sequential-Line uses two design points in the high-dimensional parameter space to define the user's search space at each iteration.  
One point is the design with the highest acquisition value, \( x^{AF} \), computed using the GP model based on the user's selections so far.  
The second point is the best candidate observed to date, \( x^{+} \).  
Connecting these two points forms a line that defines the search space.  
Users can explore along this line using a slider and select their most preferred design.  
However, searching a high-dimensional space with a single slider typically requires many iterations.  
Sequential-Line addresses this by limiting the number of parameters to six and relying on crowd-sourced data in their user study~\cite{koyama2017sequential}.

To further boost the efficiency, Sequential-Gallery was proposed later, which presents candidates in the form of a 2D gallery. 
With the gallery arranged on a 2D plane, users can explore two dimensions in each iteration. 
To define the 2D search plane, Sequential-Gallery uses $x^{+}$ as the center, $x^{AF}_{1}$ as one of the corners, and then searches for another point $x_{2}^{AF}$ as another corner, which has the highest acquisition value in the orthogonal direction to the line connecting $x^{+}$ and $x_{1}^{AF}$. 
Additionally, Sequential-Gallery defines a symmetric search space, so the 2D gallery is defined by these three points. 
Sequential-Gallery then evenly samples discrete points in this space and presents them to the users for selection.

In this work, we follow the interaction of Sequential-Gallery, extracting a 2D search space from a high-dimensional design space. 
However, instead of relying solely on the current GP to construct the search plane, we focus on leveraging prior experiences as well as a two-step BO approach to suggest design points that can boost the sample efficiency, which will be expanded in the next sub-sections.

\subsection{Transfer Acquisition Function (TAF): a Meta-Learning Method}
\label{section: TAF_introduction}

We aim to enhance PBO's performance and efficiency via meta-learning.  
Specifically, we build on the Transfer Acquisition Function (TAF)~\cite{wistuba2018scalable}, which maintains a separate model for each prior task, which is referred to as \emph{population models}.  
During deployment, TAF also constructs a \emph{current model}, a GP trained only on observations from the ongoing task.  
Here, a task refers to a user completing a preferential optimization session on an image or 3D scene with a specific target theme.  
To guide optimization, TAF aggregates acquisition values from both the population and current models using a weighted-sum approach.  
The candidate with the highest aggregated score is selected for evaluation.

We follow the prior works to use Expected Improvement (EI) as the basic acquisition function \cite{wistuba2018scalable, liao2024meta}.
EI computes the expected improvement of all candidates over the best observed objective value in the previous observations \(\mathcal{D}_{k}\), as defined by  

\begin{equation}
    \text{EI}(x, \mathcal{D}_k) = E[I(x, \mathcal{D}_k)] = E[\max(0, \mu(x) - y_{\max})]
\label{equation: expected_improvement}
\end{equation}  

where \( I(x) \) represents the predicted improvement, defined as the difference between the predicted mean \(\mu(x)\) at a sampling point \(x\) and the best objective value observed \( y_{\max} \). 
By definition, EI is the expected value of \( I(x) \), indicating the anticipated improvement over the current best observation.  
Then, the TAF with EI as the base acquisition function could be computed with:  

\begin{equation}
    a(x) = \frac{w_{M+1}E[I_{M+1}(x)] + \sum_{i = 1}^{M} w_i I_{i}(x)}{\sum_{i=1}^{M+1} w_i}
\label{equation: TAF}
\end{equation}  

\noindent
where \(M\) denotes the number of prior models, \( I_i(x) \) is the improvement at sampling point \(x\), as defined in~\autoref{equation: expected_improvement}, and \( w_i \) represents the weight assigned to the \( i \)-th prior model.

Weights determine each model's influence: higher weights yield greater impact. 
\citet{liao2024meta} employed a technique, TAF-M, that assigns weights based on prediction confidence: higher confidence yields higher weight.  
However, this approach does not adapt to the current user's feedback or ongoing observations.  
We instead adopt TAF-R, a variant that determines weights based on similarity between the current model and each population model.  
Details are provided in \autoref{sec:method}.




\subsection{Two-step Acquisition Function}
\label{section: preliminary_two_step}

To construct a 2D search plane for sequential gallery-based optimization, we require three design points. 
We compute one of these points by the two-step lookahead acquisition function~\cite{osborne2009gaussian, ginsbourger2010towards, lam2016bayesian, wu2019practical}. 
Here, we provide a short introduction to this advanced technique in BO.
Standard BO uses a one-step greedy strategy, selecting the next point based solely on the acquisition function given the current model.
In contrast, two-step BO anticipates how the next evaluation might reshape the model and impact subsequent choices. 
It does this by simulating the outcome of a potential first evaluation, updating the model hypothetically, and computing the second-step acquisition value conditioned on this simulated update.

Theoretically, this approach explicitly searches for the most effective decisions in the next round. 
Rather than merely exploiting current knowledge, two-step BO balances exploration and exploitation more effectively, making it less prone to local optima. 
Empirically, \citet{wu2019practical} demonstrates that two-step BO consistently outperforms one-step strategies across various optimization tasks.
In our specific case, since the one design point is already selected by the typical one-step greedy acquisition strategy, employing a two-step lookahead acquisition for another point complements it by encouraging more principled and forward-looking exploration. 

We build on \citet{wu2019practical}, using Monte Carlo sampling to simulate the potential outcome of evaluating a first candidate point, then updating the Gaussian Process model with this simulated result.
This is more practical as it efficiently approximates the first evaluation using posterior samples, avoiding expensive exact inference.
Using Expected Improvement (EI) as an example, the two-step acquisition function can be written as:

\begin{equation} 
    a^{\text{2-step}}(x) = EI_0(X_1) + E \left[ \max_{x_2 \in X} EI_1(x_2) \right] 
\end{equation}
\label{eq:twostep}

\noindent
$X_1$ denotes the simulated candidate(s) in the first step, $EI_0$ is the expected improvement under the current model, and $EI_1$ is computed using the updated model after the simulated evaluation.

Lastly, our approach does not compute the two-step acquisition value solely based on the current user's model. Instead, it leverages the combined predictions of all population models along with the current model, following the same aggregation principle as TAF. In other words, the two-step acquisition is also guided by TAF. We expand on this integration in the next section.



\section{\method: Meta-learning for Preferential Bayesian optimization}

We now propose \emph{\method}, a novel method designed to enable efficient \textit{preferential optimization} by leveraging prior experience from related visual design tasks.  
We define a visual design task as one in which a user adjusts the parameters controlling the visual appearance of a 2D photo or 3D scene to achieve a specific \textbf{target theme} (e.g., ``warm,'' ``vintage'').
\delete{\method operates on the Sequential-Gallery interaction, where the user is presented with a search plane and selects the most preferred design candidate at a time.
The final design is derived after several iterations of selection. }
\change{\method operates on the Sequential-Gallery interaction, which is the state-of-the-art approach for interactive PBO. 
Using this method, users are presented with a search plane and select their preferred design candidate one at a time. 
After multiple iterations of selection, the final design is determined. 
For the meta-learning component, \method adopts TAF, which maintains low computational complexity and operates effectively without the need for large datasets.}

We consider two scenarios for meta-learning.
\textbf{(1) Cross-user, same-theme}: Prior users have optimized designs for a specific theme, and the current user seeks to achieve the same theme. This setting focuses on generalizing across different users with similar goals.
\textbf{(2) Cross-user, cross-theme}: Prior users have optimized for a range of themes, while the current user is pursuing a \textit{novel, unseen} theme. This more challenging case requires Meta-PO to generalize across both users and stylistic goals.

For both scenarios, we assume a group of ``prior users'' has completed full optimization sessions --- a phase we refer to as ``population modeling.''  
The resulting population models are then leveraged to support the design process of ``new users'' during the ``deployment'' phase.  
Crucially, even users targeting the same theme may have different subjective preferences and different target images.  
Furthermore, themes themselves can vary widely in style, adding complexity to knowledge transfer.


Below, we first outline the key challenges of Meta-PO.
We then introduce the strategies and components of Meta-PO to address these challenges, followed by a detailed explanation of its workflow. 
Finally, we conduct a series of simulated experiments to evaluate the effectiveness of Meta-PO in preferential optimization tasks.

\subsection{Challenges for Meta-PO}

While \citet{liao2024meta} used TAF to improve sample efficiency in human-in-the-loop optimization, their method does not address the specific challenges of preferential optimization.
To develop \methodnospace for Sequential-Gallery-based interactions, we identify key challenges unique to preference-based tasks.

The first challenge is \textbf{how to intelligently adjust the weights of population models based on ongoing observations}.
Prior work computed weights solely from predictive uncertainty (i.e., variance) of each GP model~\cite{liao2024meta}, reflecting confidence in prior data but remaining fixed during optimization.
This is limiting for preferential tasks, where user preferences are highly subjective and vary across individuals and themes.
Static weights can cause irrelevant models to dominate. To support effective meta-learning, weights must adapt in real-time, emphasizing knowledge from prior models that align with the new user's unique preferences.

The second challenge, specific to Sequential-Gallery-based interaction, is \textbf{how to select the third point to construct a search plane}.
Constructing the 2D plane requires at least three candidates.
Following the original method, we retain the current best design (i.e., the most preferred so far) as the first point.
The second point is selected via meta-learning (TAF), which aggregates population models to identify the highest acquisition value.
However, the third point remains an open problem. The original work~\cite{koyama2020sequential} addresses this by constructing a line orthogonal to the first two points and searching along it.
While this promotes diversity, it restricts exploration to a limited subspace, possibly missing promising regions.

The final challenge is \textbf{how to flexibly balance the influence of population models and the current model over time}.
This reflects a broader trade-off: population models provide guided exploration but may suppress personalization, while the current model supports personalization but lacks prior guidance.
Though TAF internally balances these weights, the dynamics are not directly controllable or linked to practical factors like the remaining search budget.
In realistic scenarios, one may want to lean on prior models early and shift toward personalization as the budget tightens.

\subsection{Core Elements and Design Principles of Meta-PO}

\method consists of two elements: a gallery of \emph{population models} which are gathered from prior users, and a \emph{current model} which is constructed when deploying on the end-users. 
We adopted TAF, which is a weighted-sum approach to aggregate acquisition values from population models and the current model to support and guide the ongoing optimization. 

\subsubsection{Core Elements}

\textbf{Population models:}
Population models are built from the optimization experiences of prior users.
For each optimization for a specific user in population modeling, we train a GP model based on user evaluations.
Each model captures explored regions and observed outcomes, enabling effective reuse of prior knowledge.
These models are stored as a gallery and dynamically aggregated during deployment for the new users.

\textbf{Current model:}
During deployment, for each new user, we construct a dedicated GP model using only data from the ongoing task.
This model captures user-specific characteristics and evolves as new observations are gathered.

\subsubsection{Design Principles}

\textbf{To intelligently adjust the weights of population models based on ongoing observations}, we depart from prior work that assigns weights solely based on predictive uncertainty (i.e., variance) from each model~\cite{liao2024meta}. 
Instead, we use a similarity-based weighting mechanism that dynamically adapts to the current user's observed preferences. 
Specifically, we compute the similarity between the current model and each population model, assigning higher weights to those that align more closely. 
\change{The computation of model similarity will be detailed in the following section.}
This ensures that relevant prior experience exerts stronger influences, while less relevant models are appropriately downweighted, minimizing the risk of misleading the optimization.

Another principle, specific to Sequential-Gallery interaction, is the \textbf{selection of the third design point} needed to construct a 2D search plane. Prior work addresses this via a heuristic: it defines a line orthogonal to the first two selected points and searches along this line~\cite{koyama2020sequential}. However, this approach does not incorporate information from prior users and restricts exploration to a predefined subspace. In contrast, we propose a more principled strategy by combining TAF with a two-step acquisition function~\cite{wu2019practical}, which explicitly accounts for the impact of future evaluations.

The last design principle is \textbf{to enable more flexible control of the trade-off between the weights on the population models and the current model}. 
We employ a time-dependent decay factor that gradually reduces the influence of population models over the course of the optimization.
This allows population models to guide exploration in the early stages—when data from the current user is scarce—and enables the current model to take over as more personalized observations become available.

\subsection{Method}
\label{sec:method}

Following our design principles, we propose our \methodnospace. 
In this section, we first detail how the TAF is integrated into preferential optimization, and further construct a search plane (corresponding to the first two design principles). 
Then, we detail how we introduce the decay factor to dynamically balance the population models and the current model (corresponding to the third principle).

\subsubsection{Transfer Acquisition Function-Ranks (TAF-R)}

A key design principle of \method is to adaptively adjust the weights of the models based on observations. 
Rather than relying on static model confidence as in prior work~\cite{liao2024meta}, we adopt TAF-R (R stands for ``Ranks'' of model similarities), a variant that updates model weights adaptively 
\change{ based on the relative ranking of model similarities.
Model similarity is measured by how well one model's predictions align with those of another.}
\delete{by measuring the alignment between the current user and previous users.}

Specifically, TAF-R monitors all design points that the current user has compared (e.g., between \( x_1 \) and \( x_2 \)).  
For each such pair, both the current model and each population model predict which option is preferred.  
If a population model's prediction aligns with the current user model (e.g., both prefer \( x_1 \) over \( x_2 \)), its rank increases; if not, it decreases.  
These ranks are then used to compute similarity-based weights.
Finally, these weights are used to aggregate acquisition values as defined in \autoref{equation: TAF}, ensuring that population models more consistent with the user's evolving preferences exert greater influence during optimization.  
We refer readers to the original paper \cite{wistuba2018scalable} for additional implementation details.

\subsubsection{Constructing Search Plane with Two-Step Acquisition Function}

To construct the 2D search space for Sequential-Gallery, we require three distinct design points.
In addition to retaining the current best design as the first point (denoted as \( x_k^{+} \)), we must determine two additional candidates.
We utilize TAF-R to construct the second point by identifying the design with the highest acquisition value under the similarity-weighted ensemble of prior models. 
This point is denoted as $x^{AF}_{k,1}$.
While this is analogous to how Sequential-Gallery selects the second point using a single-model acquisition function, our approach leverages TAF-R to aggregate information across multiple models for more informed guidance.

To further select the third point, we integrate TAF-R with the 2-step acquisition function, as described in \autoref{eq:twostep}.
This point is denoted as $x^{AF}_{k,2}$.
This approach simulates hypothetical evaluations of the second point, updates the GP accordingly, and computes the expected improvement for all potential third candidates based on this updated model.
In the context of preferential tasks, to perform Monte Carlo sampling similar to the approach in the original paper \cite{wu2019practical}.
Specifically, we use GP to predict the mean and variance of the second query point ($x^{AF}_{k,1}$) and all points that formed the search plane in the previous iteration. 
We then sample from their predicted distributions and compare these sampled values to generate hypothetical comparison outcomes. 
These hypothetical comparisons are then used to update the GP model.
By doing so, it anticipates how the selection of the second point will influence future evaluations, allowing the third point to complement the search plane with greater diversity and long-term utility.
This strategy leads to more principled exploration than heuristic-based search.

The search plane is constructed using \( x_k^{+} \) as the center, with \( x^{AF}_{k,1} \), \( x^{AF}_{k,2} \), and their symmetric counterparts (reflected across \( x_k^{+} \)) forming the corners.
From this plane, a new set of candidate designs is sampled and presented to the user. 

\subsubsection{Balancing the Population Models and the Current Model via the Decay Factor}

To balance the influence of prior population models and the current user model, we introduce a time-dependent decay factor that adjusts their contributions throughout the deployment.  
Prior models provide useful guidance early on but may not reflect the current user's preferences.  
As more user-specific data accumulates, the current model becomes more informative.  
This motivates a dynamic weighting strategy: population models guide early exploration, while the current model dominates in later stages.
Following \citet{liao2024meta}, we define a decay function \( d(k) \) that adjusts weights based on the current iteration \( k \), parameterized by \( d_1 \) and \( d_2 \) to control decay duration and slope:

\begin{equation}
    d(k) = 
    \begin{cases} 
        1, & \text{if } k \leq d_1 \\
        1 - (k - d_1) d_2, & \text{if } d_1 < k \leq d_1 + \frac{1}{d_2} \\
        0, & \text{otherwise}
    \end{cases}
\end{equation}

\noindent
This decay factor is applied to each population model's weight: \( w_i^{+} = d(k) \cdot w_i \), allowing their influence to gradually diminish as the optimization progresses.
Thus, the overall acquisition function at iteration \( k \), denoted \( a_k(x) \), is then defined as:

\begin{equation}
    a_k(x) = (1 - d(k)) \cdot \mathbb{E}[I_s(x)] + \sum_{i=1}^M w_i^{+} \cdot I_i(x)
\end{equation}

\noindent
Here, \( \mathbb{E}[I_s(x)] \) is the expected improvement computed from the current model, and \( I_i(x) \) is the acquisition value from the \( i \)-th population model. 

\subsection{Workflow}
\label{sec:workflow}

\begin{figure*}
    \centering
    \includegraphics[width=\textwidth]{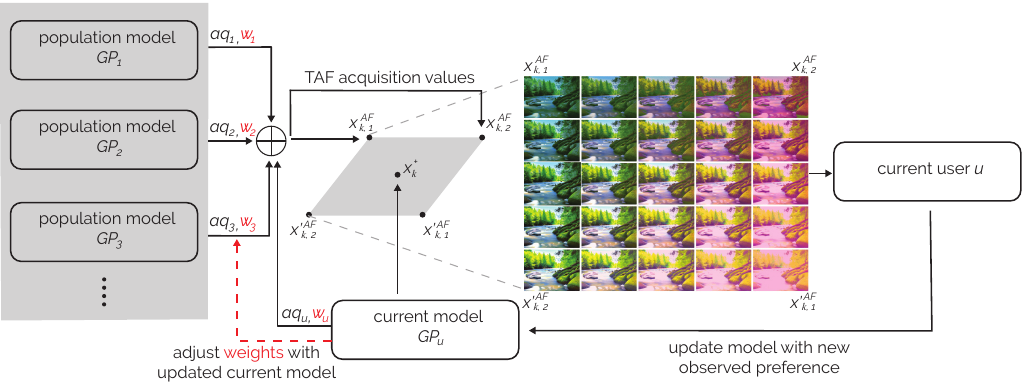}
    \caption{
        \method's working mechanism during deployment.  
        \method first stores a set of population models (shown in the gray area). During deployment, it constructs a 2D search plane per iteration using three distinct design points in a high-dimensional design space. The first point, \( x_k^{+} \), is the most preferred design so far and serves as the plane's center. The second point, \( x^{AF}_{k,1} \), is selected by aggregating acquisition values from both the population models and the current model using TAF-R. The third point, \( x^{AF}_{k,2} \), is derived from a two-step extension of TAF-R. \method then reflects these two points across \( x_k^{+} \) to create \( x'^{AF}_{k,1} \) and \( x'^{AF}_{k,2} \). Jointly, these four points define the full 2D plane; \method then samples a grid of candidates within it and presents them to the user. The user's selection updates the current model (\( GP_u \)), which in turn allows TAF-R to adjust model weights based on the similarity between \( GP_u \) and each population model \( GP_i \). A new plane is constructed in the next iteration using these updated weights and new acquisition values.
    }
    \label{fig:workflow}
\end{figure*}

The full workflow involving \method has two main phases: \textit{population modeling} and \textit{deployment}.
In the population modeling phase, prior users each complete a full preference-based optimization session using the Sequential-Gallery interaction.
Each session results in a GP model that captures the user's preference function. 
During deployment, the goal is to optimize for a new user by efficiently leveraging population models via \method.



\subsubsection{Population Modeling}

The initial search plane is constructed using three randomly selected points from the parameter space. Once the user provides the first selection from this candidate set, Meta-PO updates the GP model using this observation, as described in \autoref{eq:modelupdate}. 

In subsequent iterations, the three design points that define the search plane are determined using the acquisition function. Consider iteration \( k \) as an example. Based on all accumulated observations up to this point, denoted \( \mathcal{D}_k \), the system selects the most preferred design—i.e., the design chosen by the user in the previous iteration—as the first point, \( x_k^{+} \).

The second point is determined by the acquisition function \( a_k(\cdot) \). Specifically, the design with the highest acquisition value across the search space \( \mathcal{X} \) is selected as:

\begin{equation}
    x^{AF}_{k,1} = \arg\max_{x \in \mathcal{X}} a_k(x)
\end{equation}

To determine the third point, we employ a two-step acquisition strategy. This method anticipates the effect of evaluating the second point and selects the next best candidate under this updated belief. In the context of preferential observations, we assume the second point \( x^{AF}_{k,1} \) is preferred over all prior observations, i.e., \( x^{AF}_{k,1} \succ \mathcal{D}_k \). Under this assumption, we augment the dataset as follows:

\begin{equation}
    \mathcal{D}_k' = \mathcal{D}_k \cup \{x^{AF}_{k,1} \succ \mathcal{D}_k\}
\end{equation}

Using \( \mathcal{D}_k' \), we update the GP hyperparameters and re-estimate the similarity weights of the population models, thereby recomputing the acquisition function \( a_k'(x) \). The third point is then selected based on:

\begin{equation}
    x^{AF}_{k,2} = \arg\max_{x \in \mathcal{X}} a_k^{2-step}(x)
\end{equation}

The search plane then takes \( x_k^{+} \) as the center, with \( x^{AF}_{k,1} \), \( x^{AF}_{k,2} \), and their counterparts as corners.
The user explores the options on this plane and selects the design that best aligns with their preferences. This completes one iteration of the optimization loop. The updated preference observation is then added to \( \mathcal{D}_k \), and the three key points are recomputed to construct the next search plane.
After all population models are gathered, they are stored as a gallery to support the deployment of \method.

\subsubsection{Deployment}

When deploying \method on new users, the same overall procedure is followed. 
However, the first search plane is initialized slightly differently: only the center point ($x_k^{+}$) is sampled randomly, while the remaining two points ($x^{AF}_{1,1}$ and $x^{AF}_{1,2}$) are determined using TAF and two-step TAF. 
In this initial stage, all population models are assigned equal weights, and the current model is assigned zero weight, as no user-specific data has yet been collected.
The search plane is then constructed using the same procedure described in the population modeling phase.
In subsequent iterations, the selection of the three design points follows the same structure as in the population modeling phase. However, the acquisition functions used in deployment differ: rather than being computed solely from the current model, the acquisition function \( a_k(x) \) is now computed using TAF, which aggregates predictions from the population models based on dynamically updated weights. Similarly, the two-step acquisition function \( a_k^{2\text{-step}}(x) \) is computed using the two-step TAF formulation (\autoref{eq:twostep}).
The rest of the interaction follows the population modeling workflow: the user explores the 2D search plane, selects the most preferred design, and the model updates based on the new observation. This iterative loop continues until the optimization concludes.
The mechanism is illustrated in~\autoref{fig:workflow}.


\subsection{Simulated Tests}

We evaluated \method through simulated tests to assess its effectiveness and compare design choices (see \autoref{appendix: simulated_tests}).
We focused on two components: (1) strategies for weighting population models and (2) methods for selecting the third point to form the search plane.
We simulated user preferences by randomly shifting and scaling benchmark functions.
Non-meta baselines were used to construct population models for meta-learning variants.
All meta-learning methods improved both convergence and final performance compared to non-transfer baselines.

\change{On average, meta-learning methods required only 5–10 iterations to reach the final performance achieved by non-meta baselines.
At termination, they achieved 85.86\% lower regret compared to non-meta methods.
Among the search strategies, the two-step acquisition method consistently outperformed orthogonal exploration, reducing final regret by an average of 29.89\% across all test functions.
For model weighting, the similarity-based TAF-R approach outperformed the confidence-based TAF-M, achieving a 51.06\% improvement in average final regret.
The combination of TAF-R and two-step acquisition yielded the best overall performance and is used as the default configuration in our user studies.}

\section{Interaction}

To evaluate the effectiveness of \method in improving preferential optimization efficiency for practical applications, we implemented two applications: image enhancement and virtual scene lighting design. 
Image enhancement is a common task in everyday life, requiring users to tweak multiple parameters to achieve the desired visual outcome. 
Similarly, virtual scene lighting design involves modifying lighting parameters to control the appearance of a scene in Virtual Reality. 
Through these applications, we aim to demonstrate \method's applicability in real-world scenarios and its potential benefits for end-users. 
We conducted user studies to evaluate \method's performance in both \textbf{cross-user transfer} and \textbf{cross-theme generalization}.

\subsection{Image Enhancement}

Image enhancement is a widely used feature in social media platforms. 
We implemented an image enhancement application based on Sequential-Gallery with \method to enhance the optimization process. 
We then conducted a user study to compare the performance and user experience between \method and Sequential-Gallery.

\subsubsection{Implementation}

\begin{figure*}[t]
    \centering
    \includegraphics[width=\textwidth]{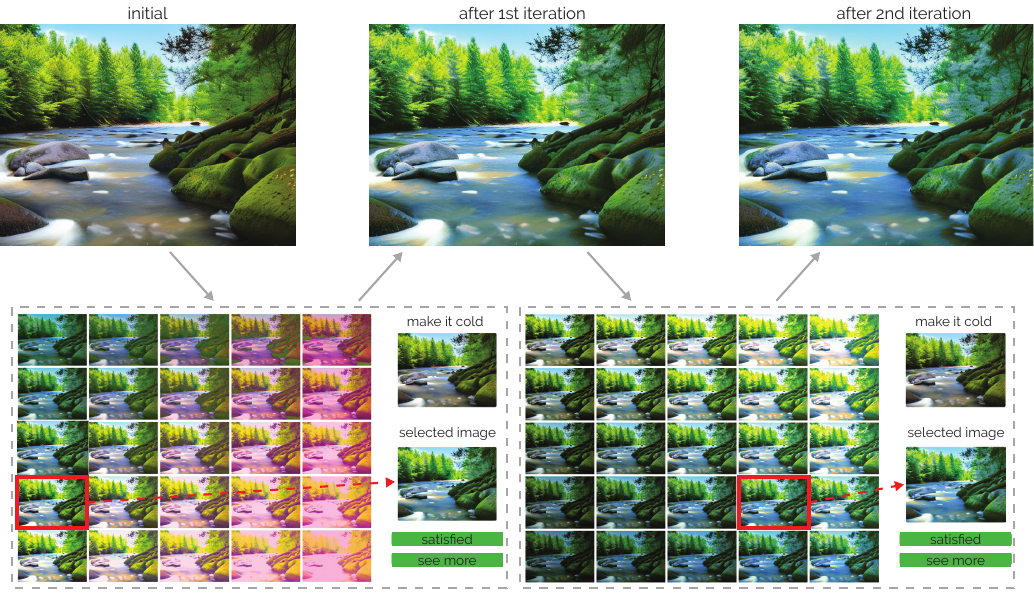}%
    \vspace{-3mm}%
    \caption{Image enhancement interface.
    Users are presented with 25 candidate images and select the one that best aligns with their desired visual effect (`theme'). 
    Our user study interface added the theme-based instruction (top right) and buttons to choose whether they were satisfied with the current iteration's outcome or needed to see additional candidates.}
    \label{fig:photo_interface}
\end{figure*}

Our application includes 12 adjustable parameters, including contrast, saturation, brightness, and color adjustments for shadows, midtones, and highlights across the RGB channels, which remain the same as used in the original Sequential-Gallery. 
\autoref{fig:photo_interface} illustrated the user interface.
In each iteration, users were presented with a gallery of candidate images and selected the one that best matched their preference. 
Upon proceeding to the next iteration, a new set of candidates was generated based on their previous selection. 
Users could terminate the optimization process and export their final design when satisfied. 
Notably, the instructions shown in \autoref{fig:photo_interface} (e.g., "make it cold") were provided for the user study but were not part of the standard user workflow.

\subsubsection{User Study Design}

We conducted a user study using our image enhancement application to investigate whether \method can effectively transfer optimization experience across both users and themes. 
To achieve this, we adopted a between-group study design with three conditions:

\begin{itemize}
    \item \notransfer{}: 
    Participants used the Sequential-Gallery interaction to optimize images without leveraging prior optimization experience. 
    This condition not only served as the baseline, but also as the participants' optimization data forming population models used in the other two conditions.
    We employed the two-step lookahead acquisition function for the Sequential-Gallery method, ensuring that the resulting models could serve as population models for subsequent conditions.
    \item \textit{Cross-users, same theme} (\crossusers{}): 
    Participants in this condition optimized images to achieve the same themes as those in the \notransfer{} condition. 
    However, their optimization process was accelerated using \method and population models provided by \notransfer{} participants. 
    By comparing results between this condition and \notransfer{}, we evaluated the effectiveness of \method in cross-user transfer.
    \item \textit{Cross-themes and cross-users} (\crossthemes{}):
    Participants in this condition optimized images to achieve different themes than those in the previous two conditions, also using \method and population models provided by \notransfer{} participants. 
    By comparing their results to \notransfer{}, we assessed the effectiveness of \method in cross-theme generalization.
\end{itemize}

We selected image themes from existing Instagram and TikTok filters.
The \notransfer{} and \crossusers{} conditions used: warm, cold, golden hour, and blue hour.
The \crossthemes{} condition used: winter, summer, pastel, and vintage.

We generated 100 landscape images using Stable Diffusion~\cite{Rombach_2022_CVPR}, all from the same prompt\footnote{“River landscape with clear water and surrounding forest.”}.
In each trial, one image was randomly selected, and participants were asked to adjust it to a target theme within 15 iterations.
In each iteration, they selected the most preferred candidate; if satisfied, they clicked “satisfactory,” otherwise “see more” to continue.
To assess optimization efficiency, we recorded the \textit{minimum number of iterations} needed to reach satisfaction.
We also collected subjective feedback using the Creativity Support Index (CSI)~\cite{cherry2014quantifying}, which measures Enjoyment, Exploration, Expressiveness, Engagement, and Satisfaction on a 7-point Likert scale (1 = strongly agree, 7 = strongly disagree).




\subsubsection{Procedure}
Participants first gave informed consent and completed a demographic questionnaire.
They were then introduced to the study and completed a warm-up trial with a randomly selected image and a given theme.
\change{Participants were given the theme names and instructed to interpret each one freely. 
The themes were presented solely as abstract terms without examples or additional guidance for two reasons: first, all themes were selected from popular social media platforms, and all participants reported prior experience using them to share photos; 
second, this approach was intended to encourage open-ended interpretation. 
This setup reflects real-world scenarios in which users aim to achieve a general visual effect without a specific target image.
}
Once ready, they proceeded to the main study.
In each trial, a random image was shown, and participants adjusted it to match a given theme.
Each trial lasted 15 iterations, regardless of satisfaction, though participants could indicate satisfaction at any point by clicking a "satisfied" button.
Each participant completed 16 trials (four themes × four images), with trial order randomized.
After all trials, participants filled out the CSI questionnaire.
The study lasted approximately 45 minutes.

\subsubsection{Participants \& Apparatus}

We recruited a total of 36 participants, with 12 participants in each group.
In \notransfer{}, participants consisted of 7 males and 5 females, aged 25–29 (\statsum{27.17}{1.59}{}). 
In \crossusers{}, the participants included 8 males and 4 females, aged 24–30 (\statsum{27.75}{1.81}{}). 
Similarly, \crossthemes{} included 7 males and 5 females, aged 25–31 (\statsum{27.92}{1.88}{}).

The study was conducted on a desktop running Windows 10, equipped with a 12th Gen Intel Core i7 CPU and an NVIDIA GeForce RTX 2080 GPU. 
Participants interacted with the user interface using a mouse and keyboard, with the interface displayed in full-screen mode on a 27-inch Philips monitor (model: 275V8LA).

We implemented both the population and current models as Gaussian Processes using the automatic relevance determination (ARD) Matérn 5/2 kernel, following the recommendation of \citeauthor{snoek2012practical}~\cite{snoek2012practical}.
The kernel hyperparameters are determined via maximum a posteriori (MAP) estimation using L-BFGS, a gradient-based local optimization method~\cite{liu1989limited}.
To find the optimal acquisition values, we employed DIRECT, a global optimization algorithm~\cite{jones1993lipschitzian}, as multiple local optima may exist.
We ran the algorithm for 50 iterations to ensure thorough exploration of the search space.
We implemented the decay factor using $d_1 = 3$ and $d_2 = 8$, based on findings from a pilot study.

\subsubsection{Results}

We analyzed the \textit{least iteration} required for achieving satisfactory and subjective feedback across conditions. 
We conducted Shapiro-Wilk tests for the \textit{least iteration} and results suggested that the normality assumption was not violated (\pvall{>}{.05}).
Therefore, we performed a one-way ANOVA to analyze the effect of \textit{Condition} on it.
For post-hoc analysis, we employed independent t-tests with Bonferroni correction.
For the qualitative feedback, we applied a Kruskal-Wallis test to each of the five metrics and we used Mann-Whitney U tests with Bonferroni correction for post-hoc pairwise comparisons.

\textbf{Meta-PO could help participants to achieve their desired image adjustments using prior optimization experience on different participants and themes.}
We first analyzed whether and when participants found their satisfied image adjustments in all three conditions.
Across all trials, participants in \notransfer{} reported achieving a satisfactory result within 15 iterations in 76.57\% of trials. 
In contrast, participants in \crossusers{} and \crossthemes{} reached a satisfactory outcome in 97.40\% and 95.83\% of trials, respectively. 
We then analyzed the \textit{least iteration} count required for participants to achieve a satisfactory result, excluding trials in which participants failed to reach a satisfactory outcome.
Statistical analysis revealed a significant difference in the \textit{least iteration} across groups (\anova{2}{33}{18.37}{<}{0.001}). 
Post-hoc comparisons indicate that \crossusers{} participants (\statsum{5.86}{1.20}{}) required significantly fewer iterations to reach a satisfactory result compared to \notransfer{} participants (\statsum{9.54}{2.19}{}, \ttest{11}{5.11}{<}{0.001}). 
This finding suggests that Meta-PO successfully accelerates the optimization process using prior experience on different human participants.
Similarly, \crossthemes{} participants (\statsum{7.41}{1.28}{}), who adjusted images to a different theme, also required fewer iterations compared to \notransfer{} participants (\ttest{11}{3.23}{<}{0.01}). 
This suggests that the optimization experience can transfer across different themes as well. 
However, post-hoc comparisons further indicate that \crossthemes{} participants required more iterations than \crossusers{} participants (\ttest{11}{-3.87}{<}{0.01}), which is expected given that the target theme for \crossthemes{} differs from the optimization experience stored in the population models.

\begin{figure*}
    \centering
    \vspace{-2mm}%
    \includegraphics[width=0.9\textwidth]{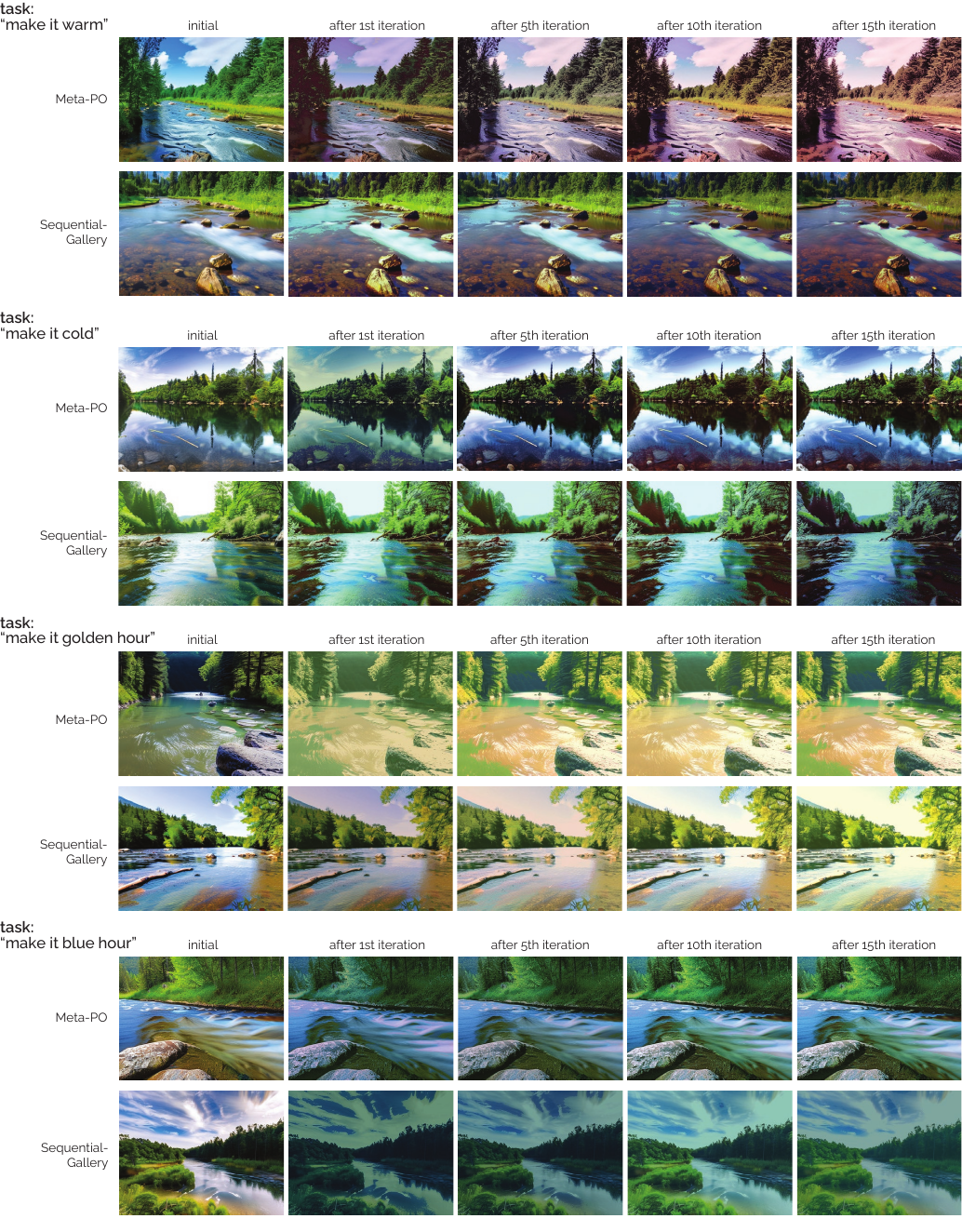}%
    \vspace{-3mm}%
    \caption{Comparison of the intermediate results generated by \method and by Sequential Gallery (baseline) across four themes.
    Participants reached satisfactory results in fewer iterations with \method than with the Sequential Gallery.}
    \label{fig:photo_results_compare}
\end{figure*}

\begin{figure*}
    \centering
    \includegraphics[width=0.9\textwidth]{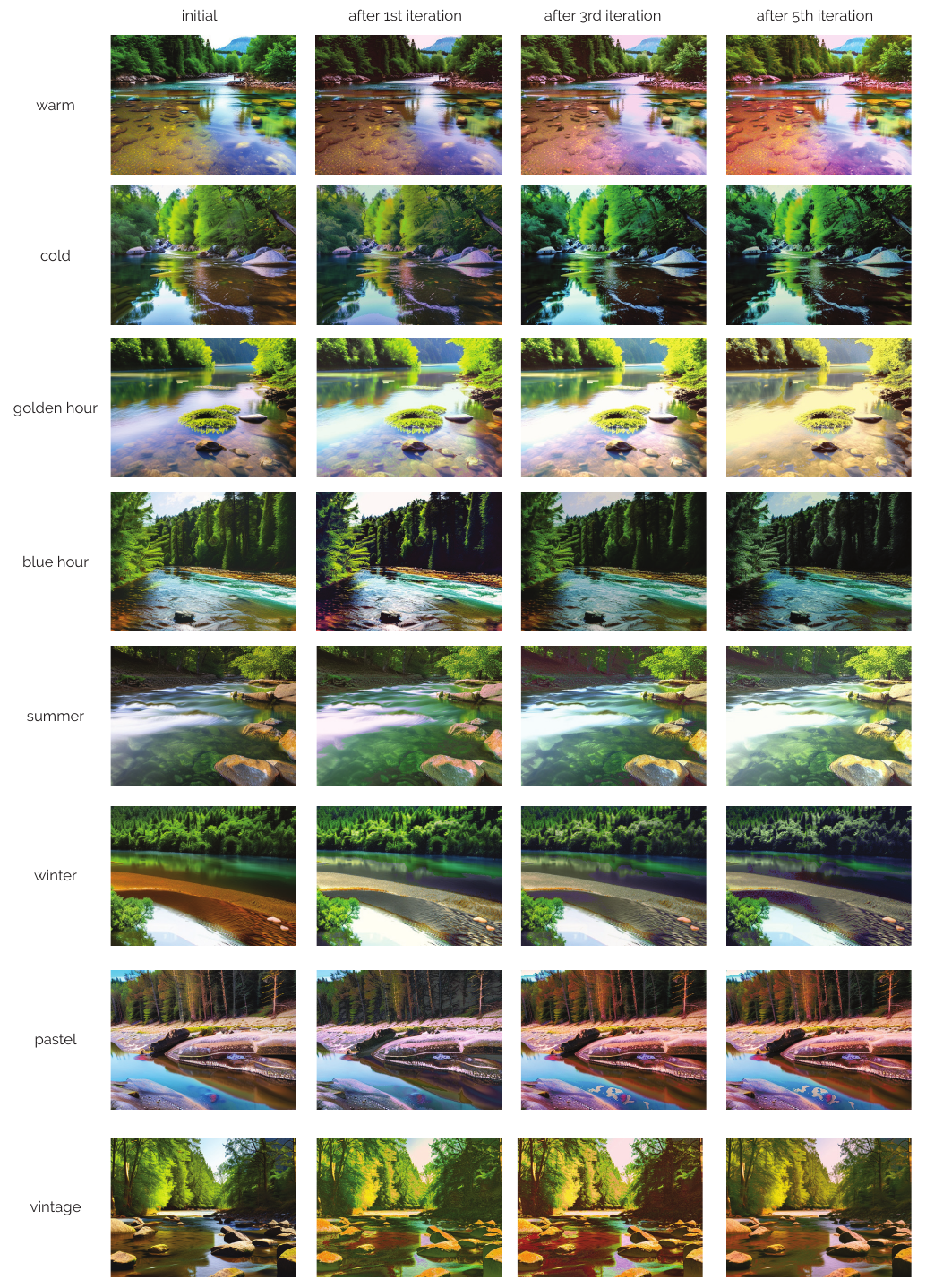}
    \caption{Intermediate results after the 1st, 3rd, and 5th iterations using \method across all eight themes. 
    Leveraging previous optimization experience, participants achieved suitable results within five iterations.}
    \label{fig:photo_results_meta_po}
\end{figure*}

The intermediate results shown in \autoref{fig:photo_results_compare} illustrate the convergence process in both \notransfer{} and \crossusers{}. 
Additionally, \autoref{fig:photo_results_meta_po} presents the intermediate results within 5 iterations for \crossusers{} and \crossthemes{}, demonstrating that participants were able to achieve satisfactory results within 5 iterations by leveraging prior optimization experience.

\begin{figure}[h]
  \centering
  \includegraphics[width=\linewidth]{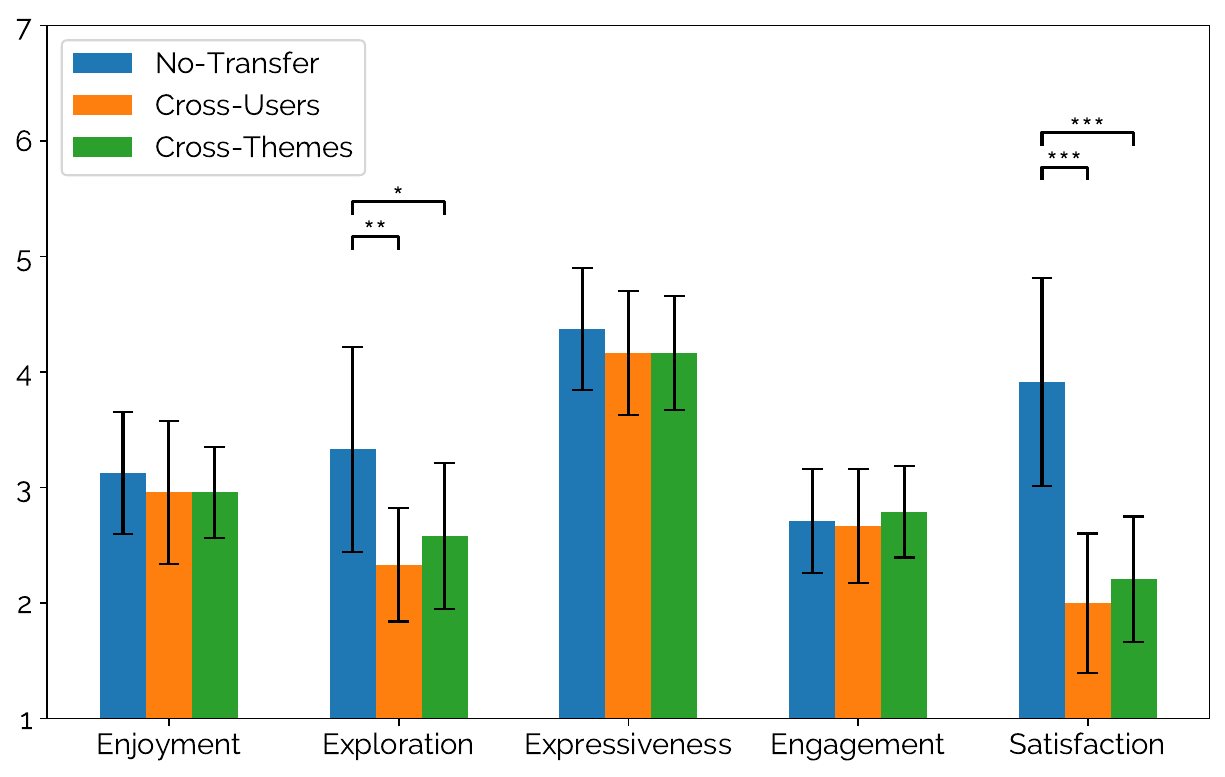}
  \caption{CSI questionnaire results for image enhancement across all three conditions (means and standard deviations). (*: $p < 0.05$, **: $p < 0.01$, ***: $p < 0.001$).}
  \label{fig: photo_subjective}
\end{figure}

\textbf{Participants perceived greater exploration potential when using Meta-PO.}
Statistical analysis revealed a significant main effect of group on \textit{Exploration} (\friedman{2}{10.53}{<}{0.01}).
Post-hoc comparisons indicate that participants in \crossusers{} (\statsum{2.33}{0.49}{}) reported a greater ability to explore diverse ideas during the optimization process compared to \notransfer{} participants (\statsum{3.33}{0.89}{}, \mann{0.83}{<}{0.01}).
Similarly, participants in \crossthemes{} (\statsum{2.58}{0.63}{}) also perceived a greater exploratory capacity compared to \notransfer{} participants (\mann{0.76}{<}{0.05}).
These findings suggest that Meta-PO enabled broader exploration during optimization, potentially due to our two-step acquisition function-based search space construction, which allows participants to explore a more diverse search space within the same number of iterations.

\textbf{Participants reported higher satisfaction with their image adjustment results when using Meta-PO.}
Statistical analysis revealed a significant main effect of group on \textit{Satisfaction} (\friedman{2}{19.77}{<}{0.001}).
Post-hoc comparisons indicate that \crossusers{} participants (\statsum{2.00}{0.60}{}) reported significantly higher satisfaction with their results compared to \notransfer{} participants (\statsum{3.92}{0.90}{}, \mann{0.94}{<}{0.001}).
Similarly, \crossthemes{} participants (\statsum{2.21}{0.54}{}) also reported greater satisfaction compared to \notransfer{} participants (\mann{0.93}{<}{0.001}).
These findings suggest that Meta-PO enabled a more efficient convergence toward participants' desired adjustments, allowing them to leverage more iterations to refine their designs, ultimately leading to more satisfactory results.

\subsubsection{Discussion}
\change{These results suggest that even partial transfer of prior experience can effectively reduce user effort and improve their experience in interactive design tasks.
By lowering the number of interactions required to reach satisfactory outcomes, Meta-PO makes PBO more practical for casual, low-effort use cases. 
This usability improvement broadens the applicability of PBO to everyday design tasks—such as color selection, font customization, and digital fabrication, where users may not have the time or expertise to engage in prolonged optimization.}

\subsection{Preliminary Study on Virtual Scene Lighting Design}


Beyond image enhancement, we also implemented a virtual scene lighting design application, allowing participants to explore the search space continuously using the joystick of the controllers. 
This application demonstrates that Meta-PO also supports more flexible exploration and interaction beyond gallery-based search. 
Similar to the image enhancement study, we conducted a preliminary user study with a total of 10 participants (across both population modeling and deployment) to demonstrate the effectiveness of \method in this use case.


\subsubsection{Design}

In this study, we included a single virtual living room scene and asked participants to design the lighting to achieve different visual themes.
This setup simulates a realistic application where users could customize their home lighting in VR before making physical adjustments.

Our study design closely follows that of the image enhancement study.
Participants were divided into two groups: \notransfer{} served as the population users, utilizing the Sequential-Gallery interaction with continuous search space without meta-learning, while \metapo{} used Meta-PO, leveraging prior optimization data from \notransfer{}.

For \notransfer{}, participants were asked to adjust the lighting to achieve warm and cold themes.
For \metapo{}, in addition to warm and cold, they also designed for golden hour and pastel lighting effects, which are different from the population participants.

We included a single light source, with participants adjusting its 3D position, RGB values, and intensity, resulting in a seven-parameter search space.
Each participant performed 15 iterations per theme and was asked to indicate the first iteration where they felt satisfied with the result.
After completing all themes, participants filled out the Creativity Support Index (CSI) questionnaire, evaluating the system across the same five key perspectives as in the previous study.

\subsubsection{Procedure}
The procedure closely follows the previous study, with one key difference: instead of selecting a preferred image from 25 candidates, participants explored the search space freely using the joystick on their controllers.
Once they were satisfied with their current design, they pressed a button on the controller to proceed to the next iteration.
During the study, a small square with a cursor was displayed in mid-air in front of the participant, illustrating the exploration space. 
This visualization helped participants better understand the boundary and identify unexplored areas.

\subsubsection{Participants \& Apparatus}

We recruited 10 participants in total, with 4 participants assigned to \notransfer{} and 6 participants to \metapo{}.
We included fewer participants in the population model for two reasons.
First, the number of parameters in this study was lower than in the previous study, resulting in a smaller search space.
We also aimed to evaluate whether Meta-PO could still effectively facilitate subsequent optimization tasks with a reduced population model.

\notransfer{} consisted of 3 males and 1 female, aged 25–29 (\statsum{27.17}{1.59}{}).
\metapo{} included 4 males and 2 females, aged 24–30 (\statsum{27.17}{1.59}{}).

Participants used a Meta Quest 3 headset connected to a desktop computer with an Intel Core i7-12700K processor, an NVIDIA GeForce GTX 2080 GPU, and 32 GB of RAM.
They interacted with the application using the Meta Quest 3 controllers.
The optimizer implementation remained consistent with the previous study.

\subsubsection{Results}
Similar to the previous study, we analyzed both the \textit{least iteration} required for achieving satisfactory and subjective feedback across groups.
\change{The primary goal of this preliminary study is to demonstrate the generalizability of \methodnospace
Therefore, we did not perform statistical analyses on the study results.}

Across all trials in both groups, every participant successfully found a satisfactory design within 15 iterations. 
This may be attributed to the lower number of adjustable parameters compared to the previous study, as 15 iterations were sufficient to 7 seven parameters.
In \notransfer{}, participants achieved their desired effect within an average of 7 iterations (\statsum{6.85}{1.48}{}). 
While participants in \metapo{}, where prior optimization experience was leveraged, reached a satisfactory result in approximately 4 iterations (\statsum{4.05}{1.39}{}) when tasked with achieving the same visual effect as \notransfer{}. 
When \metapo{} participants aimed to adjust the virtual lighting to new themes, they required an average of 5 iterations (\statsum{5.33}{1.47}{}).
These results demonstrate that \method effectively accelerates visual appearance optimization in 3D spaces, even when relying on a smaller set of prior optimization experiences from population models.

\begin{figure*}[h]
    \centering
    \includegraphics[width=\textwidth]{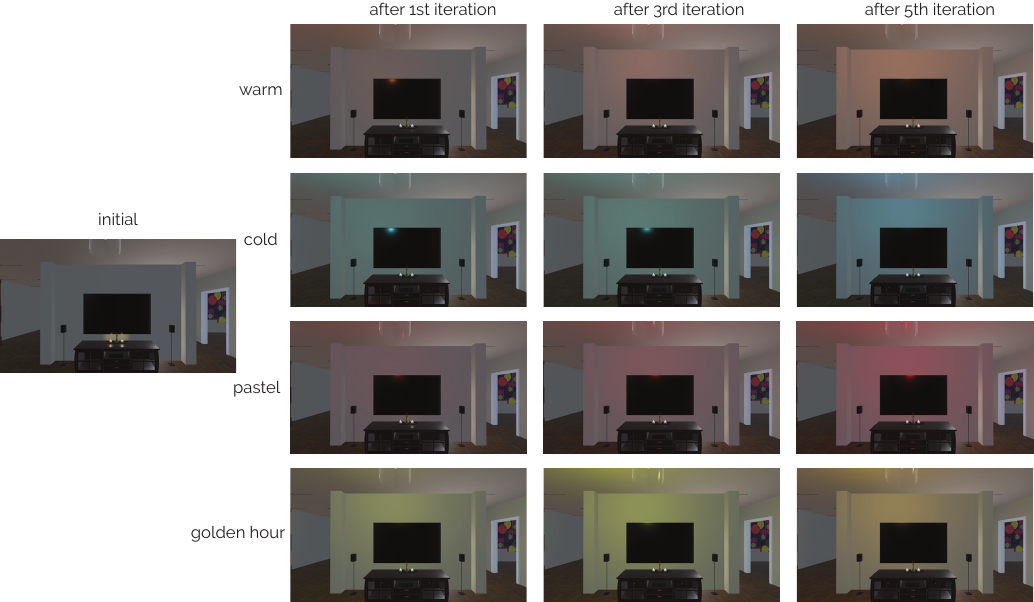}%
    \vspace{-3mm}%
    \caption{\method-based optimization of immersive lighting appearances: Intermediate results for four themes.}
    \label{fig:vr_result}
\end{figure*}

\autoref{fig:vr_result} illustrated the intermediate results using \method in all four themes.
These results demonstrated that participants could achieve satisfactory results within 5 iterations.

\begin{figure}[h]
  \centering
  \includegraphics[width=\linewidth]{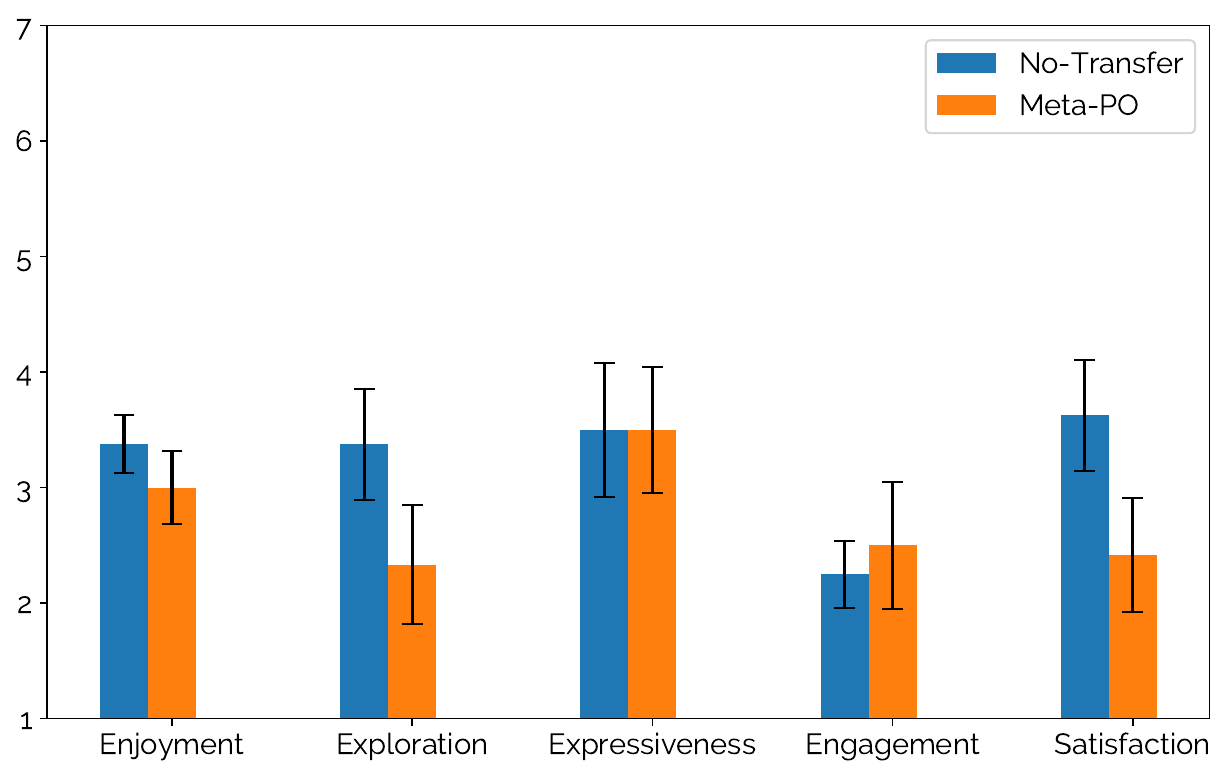}
  \caption{CSI questionnaire results for virtual lighting appearance across the two conditions (means and standard deviations).} 
  \label{fig: vr_subjective}
\end{figure}

In qualitative feedback, participants in \metapo{} (\statsum{2.33}{0.52}{}) reported feeling that they could explore a wider range of possibilities compared to those in \notransfer{} (\statsum{3.38}{0.48}{}). 
Additionally, they expressed higher satisfaction with their final lighting design (\statsum{2.42}{0.49}{}) than participants in \notransfer{} (\statsum{3.63}{0.48}{}).
These findings align with our previous study, further supporting the effectiveness of \method in enabling participants to explore more diverse design possibilities while achieving greater satisfaction with their results.

\section{Discussion}

\method accelerates the visual appearance optimization process by transferring prior users' preferences.
At its core, \method follows TAF-R to compute the similarities between the current user's specific model and all the previously gathered population models and then dynamically adjusts the weights that determine which prior model has a stronger influence to support the current optimization.
As \method builds on the interaction of Sequential-Gallery, we must derive a third design point. 
Unlike the original work's approach to searching along the orthogonal line, \method integrates the two-step acquisition function with TAF-R to identify the third point.
This simulates the potential future impact of a second design point before selecting the third, offering a broader exploration within the design space.

We evaluated \method on two visual design tasks: image enhancement and virtual lighting design.
In the image enhancement task, participants achieved their desired results in fewer than 6 iterations on average, requiring significantly fewer iterations than baseline methods that do not utilize prior experience.
Even when applying optimization experiences across different themes, participants reached their goals in under 8 iterations on average, outperforming the baseline.
Additionally, participants reported that \method allowed them to explore more possibilities and ultimately achieve more satisfying results.
In the virtual lighting design task, participants similarly found that \method enabled them to reach their desired outcomes more quickly.
These findings suggest that \method can infer users' implicit preferences and selectively transfer prior optimization experiences without requiring strong assumptions or domain-specific knowledge.
We believe \method has the potential to assist designers and users in adjusting visual appearance settings more efficiently, and to support other tasks where user preferences are central.

\subsubsection*{The impact of the number of population models}

Our study showed that \method accelerates visual appearance optimization by leveraging prior knowledge, specifically the preference models and optimization experiences of 12 participants across 192 images. 
However, we did not examine the impact of the number of population models on performance.
In principle, a larger collection of population models increases the chances of finding one or more that closely resemble the current user's preferences, thereby improving the accuracy and relevance of the guidance during early-stage optimization.
Prior work supports this idea~\cite{liao2024meta} and showed that additional models can boost sample efficiency by offering richer, more diverse prior knowledge---but the benefit gradually diminishes after a certain point.
The optimal number is highly task-dependent. 

Another key consideration is the trade-off between leveraging prior knowledge and maintaining computational efficiency.
Using more population models increases guidance but also raises computational costs at deployment time, potentially impacting real-time interaction and user experience.
This scalability issue is a shared limitation of TAF-based methods.
Future work could explore strategies such as clustering similar models or using sparse Gaussian Processes to reduce inference overhead.
More advanced directions include replacing the weighted-sum formulation with scalable Bayesian neural networks, which can leverage larger prior datasets without increasing inference cost.

\subsubsection*{Open questions for transferring experience across models and themes}


Our study demonstrated that \method enables effective knowledge transfer both across user models and across visual themes.
In particular, even when a user's target theme was unseen during population modeling, transfer was still beneficial, primarily because the new theme shared stylistic similarities with existing ones.
However, the effectiveness of transfer is highly dependent on the degree of similarity between themes and user preferences.
This raises several open questions.
Can we estimate the ``distance'' between a new theme and previously modeled themes before deployment, in order to anticipate the effectiveness of transfer?
How can we quantify similarity in a way that's meaningful for guiding preference-based optimization?
Future work should consider addressing these questions, potentially based on semantic distances and by analyzing labeled images. 
\change{Furthermore, the effectiveness of Meta-PO under entirely novel or highly abstract themes remains uncertain. 
In such cases, stylistic similarity may be ambiguous or absent altogether, limiting the relevance of prior models. 
This underscores the need for mechanisms that can detect when population knowledge is no longer applicable and adapt accordingly.}
In more extreme cases, meta-learning may not only be unhelpful; it may even hinder the optimization process.
For example, if the population models primarily capture warm-related styles, they may offer misleading suggestions when the user is targeting a cold or high-contrast theme.
Future work should explore mechanisms to detect when negative transfer is occurring and automatically fall back to optimizing from scratch, or selectively prune misleading population models based on early interaction signals.

\subsubsection*{Selecting relevant themes for initializing optimization.}
\change{In our current implementation, \method{} leverages all available prior models across themes when operating in cross-theme scenarios. 
However, as discussed earlier, incorporating models from semantically irrelevant themes may result in negative transfer, even hindering optimization performance.
A potential direction for future work is to intelligently select a subset of relevant themes to initialize the optimization. 
One potential approach is to measure semantic similarity (e.g., textual description) and visual similarity (e.g., visual features) between the current task and prior themes, and restrict initialization to those that are most relevant. This targeted selection could potentially improve convergence and enhance generalization when the model base increases.}

\subsubsection*{Alternative implementation of meta-learning for PBO}
While \method builds on TAF-R and a two-step acquisition function, alternative meta-learning mechanisms for Preferential Bayesian Optimization (PBO) could be explored.
One direction is to adopt a kernel-based approach, where meta-learning is incorporated by designing task-specific or theme-specific kernels, similar to~\citet{brochu2010bayesian}. 
Such kernels could allow smoother generalization across users or themes without maintaining an explicit set of population models.
Another direction involves Bayesian Neural Networks (BNNs), which offer amortized inference and better scalability as the number of prior tasks increases.
Additionally, one could consider meta-learned acquisition functions that adapt exploration strategies based on prior optimization behavior~\cite{volpp2019meta}, or few-shot preference modeling that learns shared representations using neural encoders.
While \method focuses on a model-based ensemble approach, future work could compare these alternatives to understand trade-offs in scalability, generalization, interpretability, and performance.

\subsubsection*{Preference Shifts During Interaction and Re-adaptation}
In practice, users' preferences may evolve during the course of optimization. As participants interact with diverse visual candidates, their goals may shift—for example, from initially seeking a “subtle warmth” to later favoring a ``dramatic vintage'' effect.
This drift presents a challenge: early observations may no longer reflect the user's current intent, leading to degraded performance if the optimizer continues to treat all past feedback equally.
To address this, future work should consider more principled models for handling non-stationary preferences.
For example, the Gaussian Process could be modified to include time-decay kernels, where older data points are downweighted as newer ones arrive --- an idea explored in non-stationary BO literature.
Alternatively, a more intuitive solution is to introduce a sliding window over the most recent iterations or use forgetting factors that diminish the influence of outdated selections.
These strategies would allow the model to stay responsive to emerging user preferences while maintaining sample efficiency.
Integrating such mechanisms into PBO could make systems like \method more robust in dynamic or exploratory design tasks.

\subsubsection*{Incorporating user ratings and rankings with preference feedback.}
\change{\method{} generates preference models by asking users to select their favorite option from a set of candidates, implicitly generating multiple pairwise comparisons between the chosen design and the others. 
While effective, future work could further increase sample efficiency by incorporating richer forms of user feedback.
For example, asking users to rank their top $K$ options would yield a richer set of pairwise comparisons. 
Alternatively, collecting scalar ratings or preference scores could provide fine-grained signals across all or multiple candidates. 
These additional forms of feedback could significantly enhance the optimizer's ability to model user preferences and accelerate convergence.
However, such extensions will also introduce increased cognitive load. Future work should carefully examine this trade-off, balancing the gains in sample efficiency against the additional effort required from users.
}

\subsubsection*{Generalization of \method to Other Applications}

Although \method is developed for visual appearance design, its underlying method, meta-learned preference-based optimization, can generalize to a wide range of human-in-the-loop design tasks.
Many domains require users to iteratively refine high-dimensional parameters solely based on subjective criteria, without an explicit or measurable objective function.
Examples include sound design, haptic interface personalization, lighting adjustment, game design, and interaction technique calibration, where user preferences drive decision-making and optimization efficiency is crucial.
In such scenarios, leveraging prior user data through population models and dynamically adapting to new preferences, as done in \methodnospace, can accelerate convergence and reduce user effort.
Future work could explore how well our \method or broader meta-learning for PBO can generalize across modalities and application contexts.

\subsubsection*{Comparison to foundation models driven by prompt iteration}

Recent advances in foundation models, such as large language and vision-language models, provide alternative means of interactive design.
Users can steer visual outcomes by iteratively editing prompts in tools like Stable Diffusion or ChatGPT, gradually approaching their desired aesthetic.
In this process, users must iteratively optimize and refine their prompts while navigating various trade-offs and requirements, such as providing explicit objective goals with verbal instructions, balancing exploration and exploitation by users themselves~\cite{xu2023compress, sabbatella2024prompt}.
By contrast, \method offers a principled, sample-efficient optimization approach that is explicitly grounded in user feedback.
It makes user preferences tractable through Gaussian Process modeling, supports transparent selection logic, and enables interpretable control over trade-offs across multiple criteria.
Users directly influence the optimization process through comparisons, not abstract linguistic cues.
Future work may explore ways to bring these benefits into prompt-based workflows; for instance, by learning preference-aware prompt templates or integrating interactive optimization loops with foundation models.

\section{Conclusion}

We introduce \method, a novel computational method that improves the efficiency of Preferential Bayesian Optimization (PBO) by leveraging prior optimization experiences through meta-learning.
It infers prior users' implicit preferences, stores them as models, and leverages them to suggest design candidates for new users. 
By dynamically weighting prior preference models based on their relevance to the current user, \method facilitates more efficient preferential optimization.
Moreover, we integrate the two-step acquisition function approach to construct a 2D search plane in a high-dimensional design space.
We evaluated the effectiveness of \method through a series of simulated tests and two user studies in both image enhancement and virtual reality lighting optimization.
Our results demonstrate that \method significantly accelerates the optimization process both in \textbf{cross-user transfer} and \textbf{cross-theme generalization}.
These results showed that \method improved the optimization efficiency on both familiar and unseen design goals.
By enabling efficient, scalable, and generalizable preference-driven optimization, \method brings personalized visual optimization closer to end-users, paving the way for intuitive and adaptive design tools that could be integrated into everyday creative workflows.
Finally, this work showcases how data-driven approaches can enable more generalizable and efficient human-in-the-loop Bayesian optimization.
We anticipate this work to shed light on the border application scope of scaling up optimization by prior knowledge across users and tasks.

\section{Open Science}

\noindent
The implementation of \method is available on our project page at \textit{\url{https://siplab.org/projects/Visual_Appearance_Optimization}}. 


\begin{acks}
Yi-Chi Liao was supported by the ETH Zurich Postdoctoral Fellowship Programme.
Zhipeng Li was partially supported by the Swiss National Science Foundation (Grant No. 10004941).
\end{acks}

\balance
\bibliographystyle{ACM-Reference-Format}
\bibliography{citations}

\appendix
\clearpage
\newpage

\section{Simulated Tests}
\label{appendix: simulated_tests}

This paper introduces \methodnospace, which leverages meta-learning to utilize optimization experience from similar past tasks, thereby accelerating the optimization process for the current task.
Importantly, our \method is built upon the interaction of Sequential-Gallery, which constructs a search plane at every iteration --- the system has to suggest three design points to form such a plane at each iteration. 

In this appendix, we conduct simulated evaluations for two primary goals. 
First, we aim to demonstrate the general effectiveness of enhancing PBO with meta-learning.
Second, two decision factors can lead to different \method implementations, including (1) different ways of computing the weights on the prior models and (2) different ways of constructing the search plane, specifically regarding the decision of the third design point.
We also aim to investigate which is the most effective combination. 
In detail, we have the following goals:

\begin{enumerate}
    \item \textbf{Validating the effectiveness of meta-learning for preference optimization:} We investigate whether meta-learning improves preference optimization by comparing our approach to a similar method that does not utilize prior optimization experience.
    \item \textbf{Comparing model weight assignment strategies:} We examine two distinct strategies for assigning weights to population models \cite{wistuba2018scalable}.
    \item \textbf{Comparing different approaches for constructing the search plane:} We specifically compare using two strategies of deriving the third decision point at every iteration, which ultimately leads to different search planes.
\end{enumerate}

As our primary goal is to compare the performance of different approaches in a meta-learning setting, we have to generate a group of similar yet different test functions.
Each specific test function can be seen as a unique user's implicit preference model. 
We simulated different user preferences by randomly shifting and scaling a given base function. 
All experiments were performed on a Windows 10 system equipped with a 12th Gen Intel(R) Core i7 CPU and an NVIDIA GeForce RTX 2080 GPU.

The results of these synthetic tests support our decision to employ TAF-R as the main approach to assigning model weights and using the 2-step acquisition function to determine the third design point.
For more details about TAF-R and the 2-step acquisition function, please refer to \autoref{sec:method}.

\subsection{Compared Methods}


Here we detail different combinations of implementing \methodnospace, which will then be compared throughout all synthetic tests.

\begin{table*}[h]
    \caption{The implementations of PBO that we compare in the simulated tests.}
    \centering
    \small
    \begin{tabular}[width=\columnwidth]{cccc}
        \toprule
        \textbf{Method Name} & \textbf{Meta-learning} & \textbf{Weight assignment strategy} & \textbf{Plane decision strategy} \\
        \midrule
        Random & No & \textbackslash & \textbackslash \\
        \midrule
        No-Transfer-O & No & \textbackslash & Orthogonal exploration \\
        \midrule
        Meta-PO-M-O & Yes & TAF-M & Orthogonal exploration \\
        \midrule
        Meta-PO-R-O & Yes & TAF-R & Orthogonal exploration \\
        \midrule
        No-Transfer-T & No & \textbackslash & Two-step acquisition function \\
        \midrule
        Meta-PO-M-T & Yes & TAF-M & Two-step acquisition function \\
        \midrule
        Meta-PO-R-T & Yes & TAF-R & Two-step acquisition function \\
        \bottomrule
    \end{tabular}
    \label{tab: ablation_methods}
\end{table*}

Following the framework of TAF \cite{wistuba2018scalable}, there exist two different approaches to assigning weights to both population models and the current model. The original work suggested two approaches, which we aim to compare.
\begin{description}
    \item [TAF-M: weight based on confidence]
    TAF-M adjusts each model's weight based on its confidence in the current sampling point.  
    This confidence is estimated using the variance of the model's predictions.  
    A higher variance indicates greater uncertainty, suggesting that the model's predictions are less reliable and should be assigned a lower weight.  
    Conversely, a lower variance implies greater confidence in the predictions, leading to a higher weight. 
    
    \item [TAF-R: weight based on prediction ranking alignment]
    TAF-R adjusts the weights based on the similarity between previous models and the current model.  
    To quantify this similarity, both the previous models and the current model generate predictions for all sampling points and rank them accordingly.  
    A previous model receives a higher weight if its ranking of predictions closely aligns with that of the current model, indicating a greater degree of similarity.  
    In contrast, if the rankings diverge significantly, the previous model's influence is reduced to prevent misleading guidance in the optimization process.  
\end{description}

We also compared two different plane decision strategies: two-step acquisition function \cite{wu2019practical} and orthogonal exploration, which is proposed in the original paper of Sequentia- Gallery \cite{koyama2020sequential}.
Note that, as introduced in~\autoref{sec:method}, we select a third point to define the search plane in addition to the current best observation, \( x_{k}^{+} \), and the point with the highest acquisition value, \( x^{AF}_{k, 1} \).  
In the below variations, the selection of the first two points ($x_{k}^{+}$ and $x^{AF}_{k, 1}$) remains the same. The only difference is the selection of the third point, which is denoted as \( x^{AF}_{k, 2} \).

\begin{description}
    \item[Two-step Acquisition Function] 
    This approach basically follows \citet{wu2019practical}. 
    It estimates the value at the design point that currently has the highest acquisition value by Monte Carlo sampling. 
    Then, it updates the GP accordingly, recomputes the acquisition values, and then selects the next point with the highest updated acquisition value as \( x^{AF}_{k, 2} \). For more details, please refer to \autoref{sec:preliminaries}.
    
    \item[Orthogonal exploration] 
    Previous work~\cite{koyama2020sequential} selected the third point by first computing the vector \( u \) formed by the current best observation, \( x_{k}^{+} \), and the point with the highest acquisition value, \( x^{AF}_{k, 1} \).  
    Then, the third point, \( x^{AF}_{k, 2} \), is determined such that the vector \( v \), formed by \( x_{k}^{+} \) and \( x^{AF}_{k, 2} \), satisfies the orthogonality condition \( u \cdot v = 0 \).  
\end{description}

Therefore, there are two ways of deriving the weights on the models and two ways of selecting the third point to form a search plane --- making four possible implementations of \method in total.

Finally, we include three baselines to compare with:  \textit{No-Transfer} refers to preferential optimization without leveraging any prior knowledge.
Yet, there are two approaches to constructing the search plane (two-step v.s. orthogonal exploration), leading to two variations. 
Additionally, we include a \textit{Random} method, where the exploration plane is constructed based on the best observation ($x^{+}$) but explored in a random direction (the plane corners are randomly selected). 

Finally, we summarize all seven methods in \autoref{tab: ablation_methods}. 
These include four meta-learning variants (two weight strategies $\times$ two plane strategies), two non-transfer baselines (No-Transfer-O, No-Transfer-T), and one random baseline. 
For the meta-learning methods, the same plane construction strategy is used in both the population modeling and deployment phases to ensure consistency. For instance, Meta-PO-M-T, which uses the two-step acquisition strategy during deployment, builds its population models using the same two-step method (i.e., No-Transfer-T). 
This alignment ensures a fair comparison and meaningful knowledge transfer across tasks.

In all simulated tests, we apply a weight decay factor with parameters $d_{1} = 5$ and $d_{2} = 0.1$.
In each iteration, we initialize the search for the highest acquisition values with 80 randomly sampled points and perform 100 iterations for each to ensure a reasonable solution.


\subsection{Benchmark Functions (Base Functions)}

To evaluate the generalizability of our method, we conducted experiments using four benchmark functions: the 3D Hartmann function~\footnote{https://www.sfu.ca/~ssurjano/hart3.html}, the 6D Hartmann function~\footnote{https://www.sfu.ca/~ssurjano/hart6.html}, Isotropic Gaussian, and Rosenbrock. 
The first two functions, Hartmann 3D and 6D, have multiple local optima, which results in a complex landscape that poses challenges for optimizers in locating the global optimum. 
The latter two illustrated that our method can be generalized to higher dimensions; we used 15-dimensional Isotropic Gaussian and 20-dimensional Rosenbrock, consistent with the setup in previous work~\cite{koyama2020sequential}.

During the optimization process, we normalize the parameter space to $[0, 1]$ and the output space to $[-1, 1]$, using regret as the optimization metric to quantify how closely the current solution approximates the global optimum.
To simulate user preferential selection, in each iteration, the design point with the highest value from the search plane will be selected as the preferred design.

\subsection{Generating Synthetic Population Models}

In the simulated tests, to generate the synthetic users that contribute to the population models, we first implement standard PBO without leveraging any prior optimization experience. 
To introduce variance across synthetic users, we shift and scale the base functions to create a diverse set of synthetic user functions and their corresponding models.
This is a common approach found in evaluating meta-Bayesian optimization~\cite{liao2024meta, snoek2012practical}.  

To shift the test function, we apply a shift factor to each dimension of the input parameter, defined as $x_n' = x_n + \delta_n$, where \( x_n \) is the original input parameter in each dimension, \( x_n' \) is the shifted input vector, and \( \delta_n \) is the shift factor, which follows a uniform distribution:  
\begin{equation}
    \delta_n \sim U(-\textit{shift\_range}, \textit{shift\_range}).
\end{equation}  

Similarly, to scale the test function, we apply a scaling factor to the function values by directly multiplying a scalar with the objective value. 
The scalar \( s \) is also sampled from a uniform distribution:  
\begin{equation}
    s \sim U(1-\textit{scale\_range}, 1+\textit{scale\_range}).
\end{equation}  

In the following tests, we introduce variations by setting \(\textit{shift\_range} = 0.05\) and \(\textit{scale\_range} = 0.1\).


\subsection{Test 1: 3D Hartmann}

We generate 20 user functions, with each optimization method running for 30 iterations per user function. 
10 of them are seen as prior users for population modeling, and the rest 10 are seen as the final target users during deployment. 




\begin{figure*}[h]
  \centering
  \includegraphics[width=\textwidth]{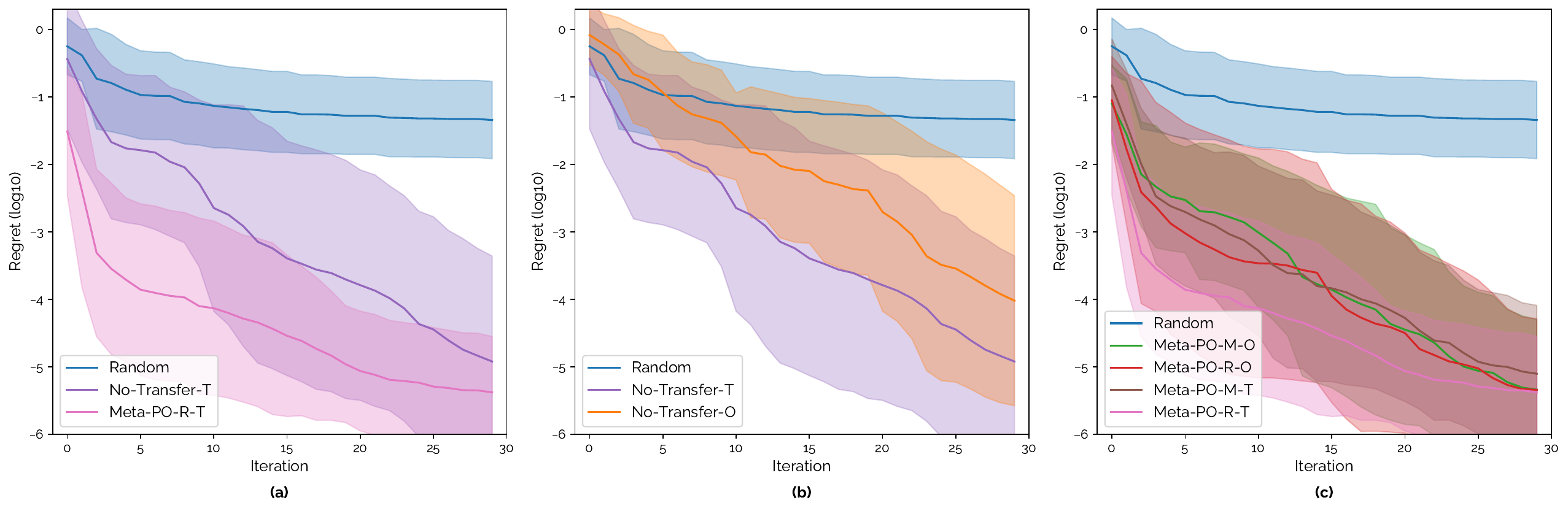}
  \caption{
        Ablation study results on 3D Hartmann. 
        The left plot compares the optimization performance of \textit{Random}, \textit{No-Transfer-T}, and \textit{Meta-PO-R-T}, showing that leveraging prior optimization experience significantly accelerates convergence and improves outcomes. 
        The middle plot contrasts \textit{Random}, \textit{No-Transfer-T}, and \textit{No-Transfer-T}, which differ in search space construction strategy, highlighting that the two-step acquisition function consistently enhances optimization efficiency. 
        The right plot compares \textit{Random} with four meta-learning-based methods, demonstrating their effectiveness in accelerating optimization, with TAF-R-based methods performing slightly better than TAF-M-based methods.
    }
  \label{fig: hartmann3d_res}
\end{figure*}

\subsubsection{Results}
\textbf{Goal 1: Validating the effectiveness of meta-learning.}
To fairly compare the performance with and without meta-learning in preferential optimization, we first examine the best performance achieved by each.
Specifically, \textit{Meta-PO-R-T} is compared against \textit{No-Transfer-T} along with \textit{Random} serving as the baseline, presented in \autoref{fig: hartmann3d_res}a.
Overall, the optimal meta-learning-based condition (\textit{Meta-PO-R-T}) achieved the overall best performance. 
While the No-Transfer preferential optimization using the two-step acquisition function (\textit{No-Transfer-T}) outperforms random exploration,  \textit{Meta-PO-R-T} achieves a regret below 0.01 in fewer than five iterations, indicating nearly converging, whereas \textit{Random} exploration requires 30 iterations to reach the same result.
Furthermore, \textit{Meta-PO-R-T} constantly yields better performance than \textit{No-Transfer-T} throughout. 
Overall, meta-learning enables faster convergence and overall better performance, showcasing its effectiveness. 

\textbf{Goal 2: Comparing model weight assignment strategies.}
We compared the two different weight assignment strategies alongside the two evaluation set selection methods (\autoref{fig: hartmann3d_res}c). 
The results indicate that meta-learning methods based on TAF-R adapt more effectively to the current test function than those based on TAF-M.
This advantage may arise from the fact that TAF-M relies on the variance of predictions at each point. 
However, when test functions are shifted and scaled, the population model's predictions may become misleading, even if they exhibit high confidence. 
In contrast, TAF-R-based methods leverage ranking alignment between the current model and population models, making them more robust to shifts and scaling in test functions.

\textbf{Goal 3: Comparing different approaches for constructing the search plane.}
We then compared the results of the two-step acquisition function with orthogonal exploration from prior work.
In \autoref{fig: hartmann3d_res}b, the results showed that while both methods constantly enable more effective exploration than \textit{Random}, the two-step acquisition function outperforms orthogonal exploration. 
Similar patterns can be found in \autoref{fig: hartmann3d_res}c where all meta-learning-based methods are compared. 
We can see that the ones involved two-step acquisition function (\textit{Meta-PO-R-T} and \textit{Meta-PO-M-T}) outperform the rest in the earlier iterations, showing more efficient convergence. 
This advantage arises from the fact that the two-step acquisition function is a more principled solution for identifying a promising candidate over the full design space rather than limiting the search space to an orthogonal line. 

\subsection{Test 2: 6D Hartmann}

Similar to the previous test, we generated 20 user functions, with each optimization method running for 30 iterations.
Among these 20 functions, 10 were seen as prior users for population modeling, and the rest 10 were seen as the final test users. 

\begin{figure*}[h]
  \centering
  \includegraphics[width=\textwidth]{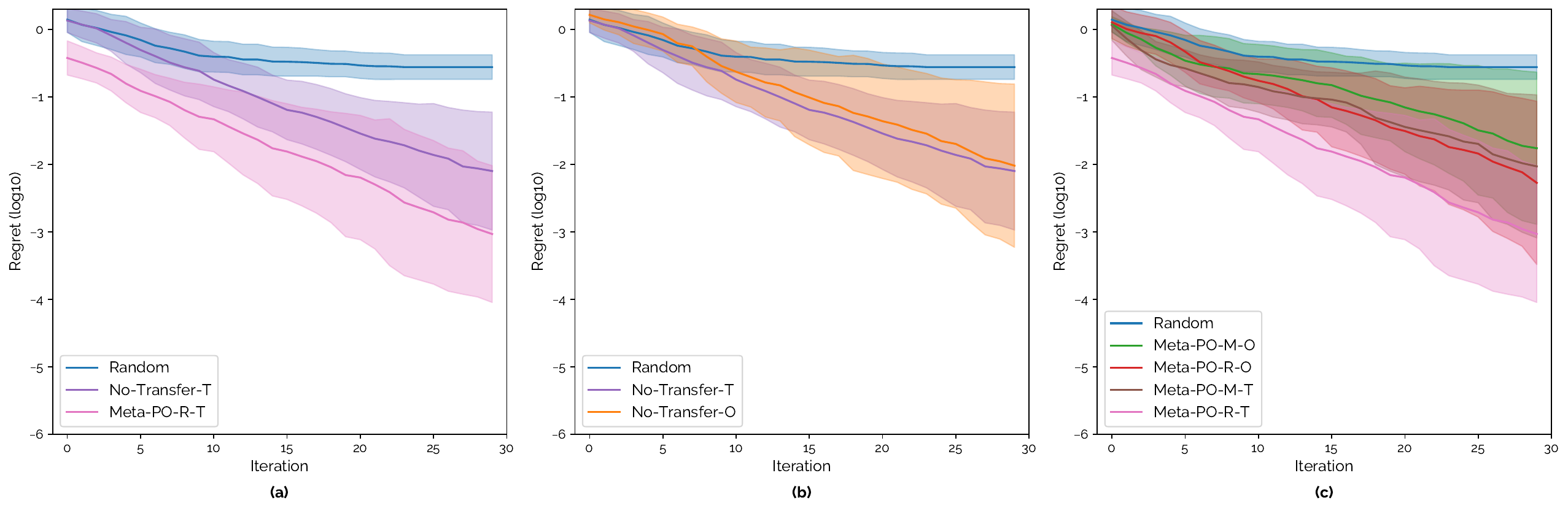}
  \caption{
        Ablation study results on 6D Hartmann.
        The main conclusions remain consistent with the previous results. 
        However, in the right plot, \textit{Meta-PO-R-T} outperforms other meta-learning-based methods, highlighting its effectiveness in handling high-dimensional optimization problems.
    }
  \label{fig: hartmann6d_res}
\end{figure*}

\textbf{Goal 1: Validating the effectiveness of meta-learning.}
Similar to the previous test, we compare the best condition with meta-learning (\textit{Meta-PO-R-T}) and without (\textit{No-Transfer-T}). 
We observe that \textit{Meta-PO-R-T} consistently achieves lower regret than \textit{No-Transfer-T} from the very first iteration (\autoref{fig: hartmann6d_res}a) and maintains superior performance throughout. 
As the dimension of the test function increases, \textit{Meta-PO-R-T} does not immediately reach a highly optimal solution but still achieves a regret below 0.1 in 10 iterations, while \textit{No-Transfer-T} achieves such a performance after 25 iterations.
These results suggest that our meta-learning-based methods can accelerate the optimization process on even more complex test functions.

\textbf{Goal 2: Comparing the weight assignment strategies.}
The results in \autoref{fig: hartmann6d_res}c suggest that TAF-R-based meta-learning methods outperform TAF-M-based methods, which is consistent with the findings from the previous simulated test. 

\textbf{Goal 3: Comparing different approaches for constructing the search plane.}
In \autoref{fig: hartmann6d_res}b, we observe that the two evaluation set selection methods achieve similar regret across all iterations. 
This could be attributed to the complexity of the 6D Hartmann test function, which has many local optima.
However, examining \autoref{fig: hartmann6d_res}b, we can observe that the two-step acquisition function (\textit{Meta-PO-R-T}) constantly results in better performance than using orthogonal line search (\textit{Meta-PO-R-O}), highlighting the benefit of using the two-step method in the meta-learning settings. 

\subsection{Test 3: 15D Isotropic Gaussian}

For the 15-dimensional Isotropic Gaussian test function, we generate 20 user functions, with each optimization method running for 50 iterations per user function. 
The increased number of iterations reflects the increased dimension of this function, requiring more iterations to converge. 
Similar to prior tests, 10 user functions are used for population modeling, and 10 are used as final test functions. 

\begin{figure*}[h]
  \centering
  \includegraphics[width=\textwidth]{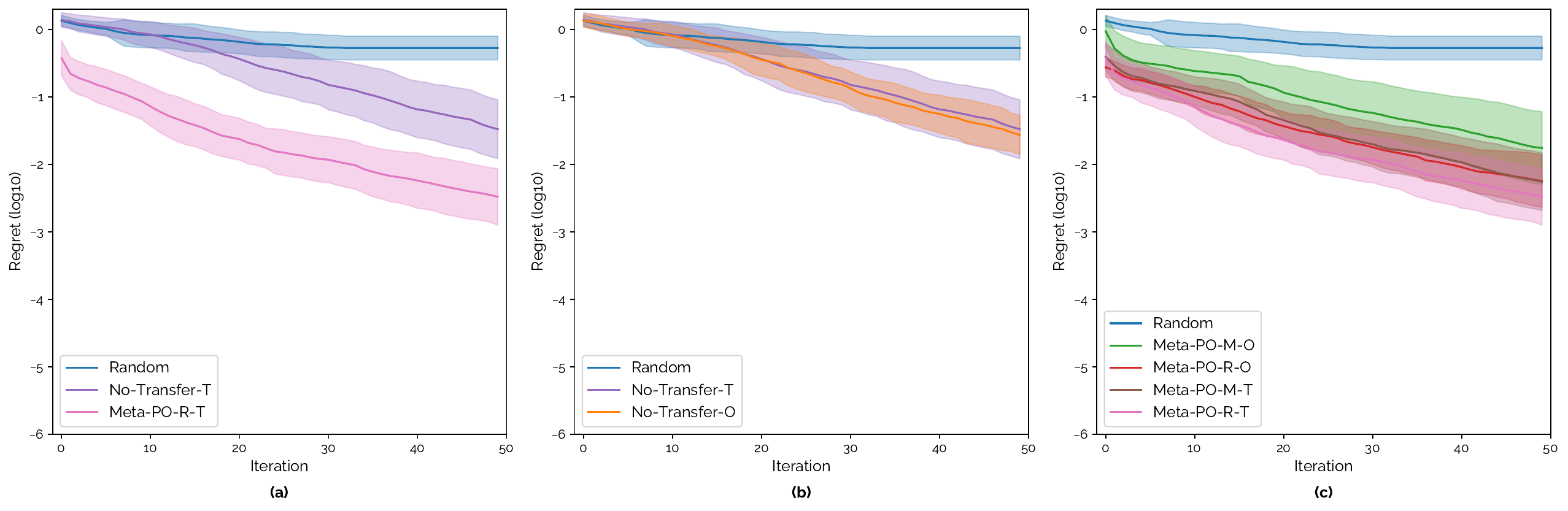}
  \caption{
        Ablation study results on 15D Isotropic Gaussian.
        The main conclusions remain consistent with the previous results. 
    }
  \label{fig: gaussian_res}
\end{figure*}

\textbf{Goal 1: Validating the effectiveness of meta-learning.}
\autoref{fig: gaussian_res}a shows that \textit{Meta-PO-R-T} achieves a regret below 0.1 within the first few iterations, whereas \textit{No-Transfer-T} requires 40 iterations to reach the same level. 
Again, the meta-learning-based approaches constantly outperform the ones without.
The \textit{Random} condition struggles to optimize the high-dimensional parameter space, though the test function is relatively naive.
This highlights the importance of acquisition functions and exploration strategies in high-dimensional optimization. 

\textbf{Goal 2: Comparing the weight assignment strategies.}
All methods with meta-learning can start from a reasonably good search point, regardless of the weight assignment strategy or evaluation selection method (\autoref{fig: gaussian_res}c).
Moreover, the results suggest that meta-learning methods based on TAF-R enable better optimization performance compared to TAF-M, which aligns with previous test results.

\textbf{Goal 3: Comparing different approaches for constructing the search plane.}
In \autoref{fig: gaussian_res}b, we observe that the \textit{No-Transfer-T} method achieves better optimization performance than \textit{No-Transfer-O} across all iterations after the initial exploration phase.
Furthermore, both \textit{No-Transfer-T} and \textit{No-Transfer-O} exhibit similar performance to the \textit{Random} method during the first 10 iterations, which is consistent with findings from prior work~\cite{koyama2020sequential}.
This is because optimization without meta-learning must first explore the high-dimensional parameter space before identifying promising regions for further exploration.

\subsection{Test 4: 20D Rosenbrock}

Similar to the previous test, we run 50 iterations on 20 user functions for each optimization method.
10 functions are used for population modeling and the rest 10 are used for final evaluation.

\begin{figure*}[h]
  \centering
  \includegraphics[width=\textwidth]{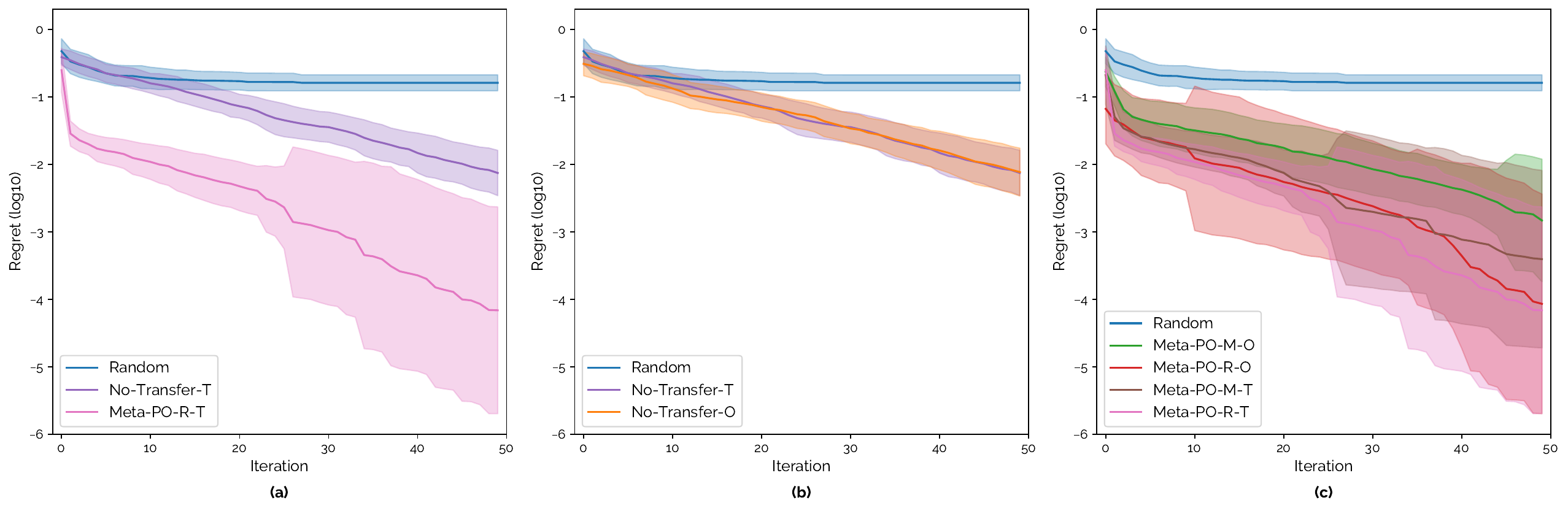}
  \caption{
        Ablation study results on 20D Rosenbrock.
        The main conclusions remain consistent with the previous results. 
        The high standard deviation observed in the results of meta-learning-based methods may be due to the log form of regret is very sensitive to small value changes, particularly when regret values are close to zero.
    }
  \label{fig: rosenbrock_res}
\end{figure*}

\textbf{Goal 1: Evaluating the effectiveness of meta-learning.}
The Rosenbrock function has a shape resembling a valley or a banana (in 2D cases). 
Its relatively simple structure makes it more manageable for the optimization process.
As a result, as shown in \autoref{fig: rosenbrock_res}a, \textit{Meta-PO-R-T} achieves a regret below 0.1 within the first few iterations by leveraging \textit{No-Transfer-T}'s experience over 50 iterations.

\textbf{Goal 2: Comparing evaluation set selection methods.}
Similar to previous test results, \autoref{fig: rosenbrock_res}b suggests that \textit{No-Transfer-T} performs comparably in most iterations and achieves a better final result than \textit{No-Transfer-O} after 50 iterations.
These findings indicate that our two-step acquisition function method outperforms orthogonal exploration on both complex test functions and high-dimensional functions.

\textbf{Goal 3: Evaluating weight assignment strategies for population models.}
\autoref{fig: rosenbrock_res}c suggests that \textit{Meta-PO-R-T} outperforms all other meta-learning methods, further demonstrating the effectiveness of combining the TAF-R weight assignment strategy with the two-step acquisition function method.
All meta-learning methods exhibit a relatively large standard deviation in the logarithm of regret after 20 iterations. 
This is because the log transformation of regret is highly sensitive to small changes in regret values, especially when the regrets are close to zero.
As a result, in some user functions, meta-learning methods can approximate the global optimum more closely, while in others, they perform less effectively, leading to a high standard deviation.

\subsection{Conclusion}

Based on the results from all four simulated tests, we conclude that meta-learning methods can accelerate the optimization process.
Particularly, TAF-R brings constantly better performance compared to TAF-M, owing to its adaptive mechanism of determining weights based on the similarities between the current model and prior population models.
Finally, the two-step acquisition function method generally results in better performance than orthogonal exploration, especially when combined with the meta-learning-based approaches. 
With these insights, we determine to employ TAF-R as the primary meta-learning mechanism with a 2-step acquisition function to conduct our user study.



\end{document}